\documentclass[a4paper,11pt]{article}
\pdfoutput=1 

\usepackage{jheppub} 
\usepackage{units}
\usepackage{graphicx} 

\usepackage{amssymb}
\usepackage{slashed}
\usepackage[normalem]{ulem}
\usepackage{bbold}

\title{
Scalable haloscopes for axion dark matter detection in the 30$\mu$eV range with RADES
}

\author{A.~\'Alvarez Melc\'on,$^a$ S.~Arguedas Cuendis,$^b$ C.~Cogollos,$^c$ A.~D\'iaz-Morcillo,$^a$ B.~D\"obrich,$^b$ J.D.~Gallego,$^d$ 
J.M. Garc\'ia Barcel\'o,$^a$ B.~Gimeno,$^e$ J.~Golm,$^{b,f}$ I.G.~Irastorza,$^g$ A.J.~Lozano-Guerrero,$^a$ C.~Malbrunot,$^b$ A.~Millar,$^{h,i}$  P.~Navarro, $^a$ 
C.~Pe\~na Garay,$^{j,k}$ J.~Redondo,$^{g,l}$ and W.~Wuensch$^{b}$}

\affiliation{$^a$ Department of Information and Communications Technologies, Technical University of Cartagena, 30203 - Murcia, Spain}
\affiliation{$^b$ European Organization for Nuclear Research (CERN), 1211 Geneva 23, Switzerland}
\affiliation{$^c$ Instituto de Ciencias del Cosmos, University of Barcelona, 08028 - Barcelona, Spain}
\affiliation{$^d$ Yebes Observatory, National Centre for Radioastronomy Technology and Geospace Applications, 19080 - Guadalajara, Spain}
\affiliation{$^e$ Instituto de F\'isica Corpuscular (IFIC), CSIC-University of Valencia, 46071 - Valencia, Spain}
\affiliation{$^f$
Institute for Optics and Quantum Electronics, Friedrich Schiller University Jena, Jena, Germany}
\affiliation{$^g$ CAPA \& Departamento de F\'isica Te\'orica, University de Zaragoza, 50009 - Zaragoza, Spain}
\affiliation{$^h$ The Oskar Klein Centre for Cosmoparticle Physics,
Department of Physics,Stockholm University, AlbaNova, 10691 Stockholm, Sweden}
\affiliation{$^i$ Nordita, KTH Royal Institute of Technology and
Stockholm
  University, Roslagstullsbacken 23, 10691 Stockholm, Sweden}

\affiliation{$^j$ I2SysBio, CSIC-University of Valencia, 46071 - Valencia, Spain}
\affiliation{$^k$ Laboratorio Subterr\'aneo de Canfranc, 22880 - Estaci\'on de Canfranc, Huesca, Spain}
\affiliation{$^l$ Max-Planck-Institut f\"ur Physik (Werner-Heisenberg-Institut), 80805 - M\"unchen, Germany }

\emailAdd{alejandro.alvarez@upct.es; sergio.arguedas.cuendis@cern.ch; cristian.cogollos@ub.edu; alejandro.diaz@upct.es; babette.dobrich@cern.ch; jd.gallego@oan.es; josemaria.gbarcelo@outlook.es; benito.gimeno@uv.es; jessica.golm@cern.ch;
Igor.Irastorza@cern.ch; antonio.lozano@upct.es; chloe.m@cern.ch; pablonm.ct.94@gmail.com; cpenya@lsc-canfranc.es; jredondo@unizar.es; Walter.Wuensch@cern.ch}

\abstract{
RADES (Relic Axion Detector Exploratory Setup) is a project with the goal of directly searching for axion dark matter above the $30 \mu$eV scale employing custom-made microwave filters in magnetic dipole fields. Currently RADES is taking data at the LHC dipole of the CAST experiment.
In the long term, the RADES cavities are envisioned to take data in the
BabyIAXO magnet. In this article we report on the modelling, building and characterisation of an optimised microwave-filter design with alternating irises that exploits maximal coupling to axions while being scalable in length without suffering from mode-mixing. We develop the mathematical formalism and theoretical study which justifies the performance of the chosen design.
We also point towards the applicability of this formalism to optimise the MADMAX dielectric haloscopes.}

\begin{document}
\maketitle
\flushbottom

\section{Introduction \label{sec:intro}}

An essential task in finding
what lies beyond the Standard Model
is to try to directly detect dark matter (DM).
Direct searches for dark matter above the MeV scale are performed by looking for the rare
recoil of dark matter against nuclei or electrons
in different target materials. A very different technique to search directly for dark matter is needed
for ultra-light dark matter. A candidate light dark matter particle is the QCD axion \cite{Wilczek:1977pj,weinberg,Peccei:1977hh,Peccei:1977ur}
which is a particle that can elegantly 
solve both the strong CP problem and constitute all of dark matter \cite{Preskill:1982cy,Abbott:1982af,Dine:1982ah}, if light enough in mass.

There are two `scenarios' for axion dark matter production leading to axions with different mass scales.
The spontaneous breaking of
the Peccei-Quinn symmetry, which gives rise to the axion as its pseudo-Goldstone-boson, takes place
only after inflation in the so-called post-inflation scenario.
On the other hand, in the so-called pre-inflation scenario, the
Peccei-Quinn phase transition happens before inflation. 
As it turns out the former
scenario generically predicts
axions with higher masses than the latter, see e.g. \cite{Irastorza:2018dyq} for a recent review.
In this article we detail on a setup that is aiming to search for post-inflationary axions, which are typically expected above $\sim 25 \mu$eV \cite{Wantz:2009it,Hiramatsu:2012gg,Buschmann:2019icd,Klaer:2017ond} using the haloscope method (which is well established at typically much lower masses).

In 1983, Pierre Sikivie suggested \cite{Sikivie:1983ip} to find axion dark matter by detecting its coupling to two photons in the following way: if a $\sim \unit[]{\mu eV}$ axion traverses a radio-frequency cavity
embedded in a magnetic field it will be converted into photons that appear as a narrow linewidth signal if the cavity is tuned to the appropriate resonant frequency.
These experiments are commonly dubbed axion haloscopes.

The haloscope technique, so far has proven efficacious in tapping into the benchmark
axion dark matter parameter space, see e.g.  \cite{DePanfilis:1987dk,Hagmann:1990tj,Du:2018uak,Braine:2019fqb, Zhong:2018rsr,Alesini:2019ajt,Alesini:2017ifp,McAllister:2017lkb,Woohyun:2016hkn}.

Many setups use solenoidal magnets and cylindrical
cavities. The diameter of the cylinder sets the frequency scale of the resonance of the fundamental mode and thus the axion mass scale which the experiment is most sensitive to. Thus probing a
high mass naively means going to small diameters.
Unfortunately, the signal power decreases as the cavity's dimension is lowered because it is both proportional to volume of the cavity and to the cavity's quality factor, which also decreases with frequency.
Thus, to explore the post-inflationary, high-mass window, novel techniques are explored.
 
One setup aiming to tackle the post-inflationary mass range of axions\footnote{Other notable efforts to search the QCD axion at `larger mass' using complementary techniques are, e.g. MADMAX \cite{Brun:2019lyf}, SIDECAR \cite{Boutan:2018uoc}, BRASS \cite{brass}, ORPHEUS \cite{Rybka:2014cya} and others \cite{Lawson:2019brd}.} is 
called RADES (`Relic Axion dark matter Exploratory Setup'). It has been recently developed
and has taken data in the magnet of the CAST experiment at CERN. Analysis of these data will be reported elsewhere \cite{CASTRADES}. Note that besides RADES, a second haloscope (CAST-CAPP) is part of the CAST physics programme, see \cite{Karuza:2018ugc,castcds} for more details.

The central idea of RADES (introduced in \cite{Melcon:2018dba}) is to develop cavity structures that can resonate at high frequencies
(above $\sim 8$~GHz) whilst not compromising on the volume of the 
cavity. Two considerations are central for this. First, the cavities should be able to search for axions in dipole magnetic fields. The reason being that dipole magnets should provide in the long term a larger magnetic volume available for axion search.
Particularly, the IAXO magnet is expected to provide a $B^2 V \gtrsim 300 {\rm T}^2 \rm{m}^3 $, and, being primarily a helioscope, would at some stage also join the haloscope searches \cite{Armengaud:2019uso}.

Secondly, as described in detail in  \cite{Melcon:2018dba}, a long
structure composed of $N$ rectangular sub-cavities, interconnected by irises,
can indeed have resonant modes with a large geometric coupling to the axion at a frequency scale that is mainly determined by the dimension of the individual cavities, see also \cite{morris,Goryachev:2017wpw,Jeong:2018ljc} for related concepts. The structure developed in \cite{Melcon:2018dba} achieves maximum coupling to the axion on the fundamental mode. However, the problem of this first approach is that mode mixing effects become very severe as the order of the structure ($N$) increases because as $N$ increases the modes tend to clutter at the edges of the filter bandwidth. 

The present article describes and reports on the testing of a novel concept that has been recently developed and tested for RADES.

In this new microwave filter design, we design the {\it central} mode to couple maximally to the axion. 
The resulting structure thus largely mitigates the mode mixing effect present in the previous approach \cite{Melcon:2018dba}. 
It is shown that this can be achieved by building a filter with
alternating inductive and capacitive irises.
The theoretical modelling of such a filter will be described in section~\ref{sec:alternating_theory}, including an application to dielectric haloscopes such as MADMAX.
The filter production and performance will be described in section~\ref{sec:alternating_experiment}. We conclude and give future prospects in section~\ref{sec:conclusions}.

\section{\label{sec:alternating_theory} 
Analysis of the cavity filter with alternating irises }

The first RADES prototype built and tested in 2017/2018 (described extensively in \cite{Melcon:2018dba})
was a $\sim15$cm long H-plane filter consisting of $N=5$ sub-cavities connected through inductive irises. 
This structure was designed to fully couple its first (lowest frequency) resonant mode to the axion field, but it is not optimal when going to a large number of sub-cavities because of the larger mode-mixing among adjacent resonant modes of the filter. 
We have solved this problem by designing an optimal filter with 
minimal mode-mixing of the resonant mode to the axion field, which consists on a cascade of cavities connected by alternating 
 inductive (H-plane) and capacitive (E-plane) irises. 
We also show by simulations that the new design is robust to mechanical tolerances.

\subsection{Theoretical model for a cavity with alternating irises \label{sec:theory_model}}

The general formalism that models the coupling  of a microwave filter with $N$ subcavities to the dark matter axion field 
was introduced in the Appendix of  \cite{Melcon:2018dba}. There, we demonstrated that the frequency-domain Maxwell equations 
in the background of the axion DM field\footnote{We use natural Units $\hbar=c=k_B=1$ with Lorentz-Heaviside convention.} led to the matrix equation
\begin{equation}
\label{eq:formalism}
(\omega^2 \mathbb{1} - \mathbb{M}) \vec{E} = -g_{A\gamma} \, B_e \, A_0 \, \omega^2 \, \vec{\cal G} \; ,
\end{equation}
where $\vec{E}$ is an $N$-dimensional column vector containing the electric field amplitudes of the fundamental mode
of each sub-cavity, $B_e$ is the static external magnetic field (here taken to be homogeneous), $g_{A\gamma}$ is the axion-photon coupling, and $\mathbb{1}$ is the unitary matrix of order $N$. The electric fields are sourced by the axion DM field $A$, which is taken to be homogeneous and harmonic $A=A_0 \, e^{-j \omega t}$, $j$ being the imaginary unit $j \equiv \sqrt{-1}$. The frequency $\omega$ is dominated by the axion 
mass $m_a$ such that $\omega=m_a+{\cal O}(10^{-6}m_a)$. The energy density of the DM halo is given by the harmonic oscillator result $\sim |A_0|^2m_a^2$ ($\sim 0.45$ GeV/cm$^3$ around the position of the Earth). The symbol $\vec{\cal G}$ represents a vector of form factors describing the coupling of the axion to the 
individual cavities. Its most relevant feature is that since the coherence of the axion field is very long, all elements of the vector are \emph{in phase}. For the case of our interest where all subcavities are (almost) the same in size, 
\begin{equation}
\label{eq:vecG}
\vec{\cal G} = {\cal G}_0 ~ (1,1,1,...,1) \ .
\end{equation}
 The most important element in Eq.~(\ref{eq:formalism}) for the cavity design is $\mathbb{M}$, which constitutes the $N\times N$ complex matrix encoding the properties of our filter.
By connecting the cavity sequentially by irises only to nearest-neighbours the matrix becomes tridiagonal
\begin{equation}
\label{eq:matrixexplicit}
\mathbb{M} = 
\left(
\begin{array}{c c c c c c}
{\tilde \Omega}_{1}^{2} & K_{12} & 0 & 0 & 0 & 0 \\
K_{21} & {\tilde \Omega}_{2}^{2} & K_{23} & 0 & 0 & 0\\
0 & K_{32} & {\tilde \Omega}_{3}^{2} & K_{34} & 0 & 0 \\
0 & 0 & \ddots & \ddots & \ddots & 0 \\
0 & 0 & 0 & \ddots & \ddots & \ddots \\
0 & 0 & 0 & 0 & K_{N,N-1} & {\tilde \Omega}_{N}^{2} \\
\end{array}
\right) . 
\end{equation}
The diagonal entries are ${\tilde \Omega}_q^{2} = \Omega_q^{2} - j \omega\Gamma_q$, where $\Omega_{q}$ are the eigenfrequencies of the $q$-th individual cavity considered in isolation 
and $\Gamma_q$ their loss rates due to the coupling-port and ohmic losses (considered small). The off-diagonal elements, $K_{q+1,q}$,$K_{q,q+1}$ parametrise the couplings between adjacent cavities.  
The characteristic modes of the filter cavity are the eigenvectors of the matrix $\mathbb{M}$, denoted as $\vec{e}_i, i=1,...,N$, corresponding to eigenfrequencies\footnote{Note that we use capital $\Omega$ for eigenfrequencies of individual cavities and we will use lowercase $\omega_i$'s to denote the eigenfrequencies of the entire filter, but we use $\Gamma_q$ or $\Gamma_i$ to denote the loss rate in a cavity or the whole mode, respectively. } $\tilde \omega_i^2=\omega_i^2 - j \omega \Gamma_i$.  
 
We read out the cavity through a port in the $q_r$-th cavity, typically the first. The power, expected to be tiny, is proportional to the E-field in that cavity, 
which itself can be expressed as a sum over the fundamental modes of the filter, 
\begin{equation}
(\vec{E})_{q_r} \simeq g_{A \gamma} \, B_e \, A_0 \, \omega^2 \, \sum_{i}  \left( \vec{e}_i \right)_{q_r} 
\, \frac{\vec{e}_i \cdot \vec{\cal G}}{\omega^2-\omega_i^2+j \omega \Gamma_i}  
\end{equation}
To optimally search for an axion, one should construct a cavity that has one of its eigenvectors ideally aligned to the excitation (\ref{eq:vecG}), i.e. 
$\vec{e}_1 = (1,1,1,...,1)$. Having a large amount of freedom to design $\mathbb{M}$ there are potentially many possible solutions, so we could in principle add more constraints to our design, like having the largest quality factor possible, better separation between modes, etc. 

Our first prototype formed by inductive cavities \cite{Melcon:2018dba} fulfills just the alignment with $\vec{\cal G}$ requirement for the first resonant mode, i.e. $\vec{e}_1 \propto \vec{\cal G}$. It might seem that using the lowest frequency mode is advantageous as it is very easy to recognise, but unfortunately it is not easy to isolate in the large $N$ limit. This can be easily seen in the simple model of a fully symmetric inductive filter having all equal cavities and irises, i.e. $\widetilde \Omega^2_1=... =\widetilde \Omega^2_N=\widetilde \Omega^2_q$ and $K_{12}=K_{N-1,N}=K<0$ (note that our first prototype is just a small perturbation of this model). The eigenvalues of the resulting Toeplitz matrix are well known,  
\begin{equation}
\label{eq:general_eig}
{\tilde \omega}^2_i = \widetilde \Omega^2_q + 2 K \cos \left(\frac{i\pi}{N+1}\right) \quad ; \quad i=1,...,N .   
\end{equation}

The mode coupling the most with the axion for $K<0$ (inductive coupling) is the fundamental  ($i=1$), as in this situation its eigenvector has all positive components, i.e. the electric fields of the cavities oscillate
in phase.

One undesirable feature of using the lowest frequency mode is the fact that the distance to the next mode decreases very fast with increasing $N$, 
\begin{equation}
\label{eq:simple}
\omega_1 - \omega_2 \sim \frac{K}{\Omega_q} \, \left( \cos\left(\frac{\pi}{N+1}\right) - \cos\left(\frac{2\pi}{N+1}\right) \right) \sim \frac{K}{\Omega_q} \, \frac{3\pi^2}{2 (N+1)^2}  \, \, , \text{  for  }  \,      N >> 1  
\end{equation}
where ohmic losses have been neglected. However, an imperfect tuning system or simply small imperfections in the manufacturing of the cavity are going to shift the values of the eigenfrequencies in a small but unpredictable way.  The above calculation shows that using a large number ($N$) of coupled cavities makes it hard to unambiguously identify the resonant mode of the axion. 
In that case, when two modes are very close in frequency, a very small perturbation (a change of the input-port coupling or a mechanical tolerance error) can have a large impact on the mode mixing of the microwave structure. 
These issues lie against the obvious observation that we would ideally use as many cavities as possible because the 
total signal power scales as $N$. 

A second undesirable aspect of the lowest frequency mode is that it is itself expected to be more sensitive to tolerance errors in the manufacturing process than central ones. In the zero $K$ limit all cavities would resonate at frequency $\Omega_q$ and by coupling them we split the eigenfrequencies around this value. Errors in the manufacturing process of the cavities dimensions and couplings add up in this mode with the same sign, while in other modes some degree of cancellation is expected.  The simplest example is an error in $K$ to $K+\delta K$ which shifts $\omega_1$ into $\sim \omega_1 + 2 \, \delta K$, and leaves $\omega_{N/2}$ almost unchanged.

Added together, these undesirable aspects of the simplest cavity-coupling concept prevent us to build cavities with $N$ much larger than $30$ or so. With the goal of searching for the ultimate sensitivity of this technique we have looked for improved technical solutions. 
Our basic idea stems from the observation that mode crowding is at the root of the above ``evils'', and that \emph{the central mode} of the simplest filter studied above is the best separated from the rest in the large-$N$ limit. Indeed, we find that the distance to its neighbours scales only as $1/N$,  
\begin{equation}
\omega_{N/2}-\omega_{(N/2)+1} \sim \frac{K}{\Omega_q} \, \frac{\pi^2}{2 (N+1)}   \, \, , \text{  for  } \,   N \gg 1
\end{equation}
and has most likely the largest degree of cancellations under perturbations of the original design.  
For these reasons, we decided to seek for a filter matrix with the axion-coupling eigenvector as its central mode.

One such possibility is given by choosing
\begin{equation}
\label{OMt}
\mathbb{M} = 
\omega_0^2
\left(
\begin{array}{c c c c c c}
1-k & k & 0 & 0 & 0 & 0 \\
k & 1 & -k & 0 & 0 & 0\\
0 & -k & 1 & k & 0 & 0 \\
0 & 0 & k & 1 & -k & 0 \\
0 & 0 & 0 & \ddots & \ddots & \ddots \\
0 & 0 & 0 & 0 & -k & 1 +k \\
\end{array}
\right)  ,  
\end{equation}
where an arbitrary frequency $\omega_0^2$ has been factored out and $k=K/\Omega_q^2$ is a dimensionless coupling.  It can be  easily seen that indeed the matrix in (\ref{OMt}) has an eigenvector $\vec{c} = (1,1,1,...,1)$ with eigenvalue $\omega^2_0$, expected to be in the centre of the modes split by the coupling.  
Actually, for odd $N$ this is exactly the central mode and has the largest distance to the nearest neighbours. 
The frequencies of all the cavities are the same except the first and last, which are corrected by $1-k$ 
or $1+k$ depending on the sign of the coupling with the adjacent cavity.  Similar corrections were required in \cite{Melcon:2018dba} to modify the eigenmode 
$\vec{e}_1$ of (\ref{eq:matrixexplicit}) to become $\propto (1,1,1,...,1)$. 
But the new key ingredient is the introduction of alternating signs in the coupling between adjacent cavities. 
A priori it is not clear that designing a physical filter with these characteristics is possible. However, as we will see, the trick to achieve it is to use an alternation of inductive and capacitive irises, which provide negative and positive couplings, respectively.

\subsection{Some theoretical remarks}
\label{s:theo_remarks}

Let us motivate a little more the form of the coupling matrix with alternating irises. 
It turns out to be simple to develop from the simplest tridiagonal Toeplitz case \eqref{eq:matrixexplicit} with all individual eigenfrequencies equal and normalised to $1$ and all couplings equal and given by $k$. 
We call this simple tridiagonal Toeplitz matrix $\mathbb{M}_S$. 
Once more, we recap the eigenvalues of its $m$-th mode, 
\begin{equation}
\lambda_m = 1+2 k \cos \beta_m \quad, \quad \beta_m = \frac{\pi m}{N+1} \ . \end{equation}
while the filter eigenmodes, $\mathbb{M}_S \bar e_m = \lambda_m  \bar e_m$,  have components given by
\begin{equation}
(e_m)_q = \frac{\sin (\beta_m q)}{\sqrt{(N+1)/2}}. 
\end{equation}
Now we want to design a new filter, given by a new matrix $\mathbb{M}$, such that the mode that couples the most to the axion, is the most separated from neighbours. The freedom is such that we start by seeking for a solution close to our experience. We decide to keep the same eigenvalues as that of $\mathbb{M}_S$ and aim at coupling the axion to the central mode, which we have already identified as the one with largest distance to neighbours. For an odd number $N$ of cavities the central mode is characterised by $c=(N+1)/2$ and for an even number of cavities it can be either $c=N/2$ or $c=N/2+1$.   

Now note that if $\bar e_m$ is an eigenvector of $\mathbb{M}_S$ then 
$P{\bar e}_m$ is an eigenvector of $\mathbb{M}=P\mathbb{M}_S P^{-1}$ with the same eigenvalue. 
Here $P$ is any suitable matrix.  
Therefore we want to identify a transformation $P$ such that it takes the central mode (the one with furthermost neighbours) into one that couples the most to the axion, $P \bar e_{c} = \bar e_{1}$ or ideally even $\propto (1,...,1)$. However, doing so blindly we might end up with matrices $\mathbb{M}$ that are not tri-diagonal and thus do not correspond to a linear filter where only couplings to neighbours are allowed (and are thus suitable for long accelerator magnets). 
The simplest solution to couple $P {\bar e}_c$  to the axion is to make $P$ diagonal with entries, $+1,-1$ so that the signs of $P$ compensate any sign change of ${\bar e}_c$, $P {\bar e}_c$ has all entries either positive or negative and thus represents a mode with all cavities in phase.  
Now, the key observation is that the central mode of $\mathbb{M}_S$  has components that \emph{alternate sign every two cavities} because 
$\beta_c = \pi/2 + {\cal O}(1/N)$. 
The case where $N$ is even is ambiguous and forces us to choose what we refer as the central mode, so we choose $c=N/2+1$. 
Thus, for the central modes a clever and simple choice for $P$ is\footnote{If we choose $c=N/2$ for even-$N$ the clever choice to couple to the axion is $P=\text{diag}\{1,1,-1,-1,..\}$, i.e. also alternates signs every two cavities but the first change happens between cavities 2 and 3 and not 1 and 2.}
\begin{equation}
    P = \text{diag}\{1,-1,-1,1,1,-1-1,...\}.
\end{equation}
This simple choice makes our construction very explicit. 
First, we can compute the new coupling matrix, which turns out to have alternating signs in the diagonals, 
\begin{equation}
    \mathbb{M}=P\mathbb{M}_SP^{-1} = 
    \left(
\begin{array}{c c c c c c}
1 & -k & 0 & 0 & 0 & 0 \\
-k & 1 & k & 0 & 0 & 0\\
0 & k & 1 & -k & 0 & 0 \\
0 & 0 & -k & 1 & k & 0 \\
0 & 0 & 0 & \ddots & \ddots & \ddots \\
0 & 0 & 0 & 0 & k & 1 \\
\end{array}
\right)  \equiv \mathbb{M}_W   ,   
\end{equation} 
which is essentially equal to the proposed~\eqref{OMt}, the only difference being the optimisation of the $[M]_{11}$ and $[M]_{NN}$ elements. 
Second, all the eigenvectors of $\mathbb{M}_W$, $\mathbb{M}_W \bar w_m = \lambda_m \bar w_m$,  follow from those of $ \mathbb{M}_S$ by reweighing with the alternating signs of the $P$ diagonal, $\bar w_m=P \bar e_m$, 
\begin{equation}
\label{Wmodes0}
(w_m)_q = (-1)^{s(q)}\frac{\sin (\beta_m q)}{\sqrt{(N+1)/2}}\quad, \quad 
s(q) = \text{floor} \left( \frac{q}{2} \right).
\end{equation}

Third, by construction the eigenvalues are \emph{exactly the same} than those of the base design $\mathbb{M}_S$. 
Finally, we can compute analytically the overlap of the central mode with the axion mode, 
\begin{equation}
    \frac{1}{\sqrt{N}}(1,..,1)\cdot \bar w_c = \frac{1}{\sqrt{N(N+1)/2}}
    \sum_q (-1)^s\sin(\beta_cq) 
    \xrightarrow{N\to \infty} 
    \begin{cases}
    1/\sqrt{2} & (N-\text{odd})\\
    \sim 0.900 & (N-\text{even})\\
    \end{cases},\label{eqn:altmodes}
\end{equation}
which is already quite good in the $N$-even case where none of the components of the central mode is zero. In the odd case, half of the cavities have zero E-field (the even-ones), which explains the relatively small asymptotic value.  

As a bonus, this simple construction inspires a straight-forward way of obtaining a central mode which is optimised to couple to the axion, i.e. which is proportional to $(1,1,...,1)$. The method is to choose a diagonal matrix $P$ such that its diagonal elements are the inverses of the elements of the eigenvector with $m=c$, i.e. $[P]_{qq}=(e_{c})^{-1}_q/\sqrt{N}$. With such a construction, $\bar w_c = P \bar e_c =(1,1,...,1)/\sqrt{N}$, as we desire. This method has a couple of disadvantages: it does not work for odd values of $N$ (because $\vec e_c$ has some zeros) and produces coupling matrices $\mathbb{M}$ which have entries of slightly different values. Our proposal \eqref{OMt} has the great virtue of having all equal cavities and bi-periodic couplings (except for the 1st and last) and works for $N$ even and odd. 

\subsection{Effects of finite tolerances}

We can use our simple and analytical construction to calculate the sensitivity of the filter to tolerances. We can compare the sensitivity of the simple filter with $\mathbb{M}_S$ and the alternating structure with $\mathbb{M}_W$. We will assume that the final optimisation to make the first or central mode to couple perfectly to the axion does not affect qualitatively our results. 

Consider that, due to manufacturing errors, 
our simple filter is described by a coupling matrix  $\mathbb{M}=\mathbb{M}_S+\delta\mathbb{M}$. 
We assume that the errors will induce deviations in the design values of the cavities eigenfrequencies and the couplings, i.e. $\delta\Omega_q^2, \delta K_{q,q+1}$, so the error matrix $\delta \mathbb{M}$ will be tridiagonal, symmetric and will have $N+N-1=2N-1$ a priori unknown parameters (we neglect effects of losses here because they are smaller). 
We will assume for simplicity that the parameters are uncorrelated, although this will generally not be the case because some manufacturing errors will be systematic.   
At first order in the perturbation, the eigenvalues are shifted by 
\begin{equation}
    \delta \lambda_m = (\bar e_m)^T \delta \mathbb{M}(\bar e_m) .
\end{equation}
If we assume Gaussian errors in the parameters of $\mathbb{M}$, the expectation value of the shift is zero.  The variance can be easily computed under the same assumptions, 
\begin{equation}
    \langle(\delta \lambda_m)^2\rangle = 
    \sum_{q=1}^N 
    \langle(\delta \Omega_q^2)^2\rangle (e_m)_q^4
    +\sum_{q=1}^{N-1}
    \langle(\delta K_{q,q+1})^2\rangle (e_m)_q^2(e_m)_{q+1}^2
\end{equation}

This estimate of the variance of the eigenfrequencies is exactly the same for our alternating filter matrix\footnote{Although the results would in general differ if we take into account correlations between $\delta\Omega^2$'s and $\delta K$'s. } $\mathbb{M}_W$. 

Therefore we can compare the distortion of the mode $m=1$ that we would use in the simple case with the alternating structure for which $m=c \simeq N/2+1$. In the $N\to \infty$ limit and assuming all diagonal and off-diagonal variances to be equal we find, 
\begin{eqnarray}
    \langle(\delta \lambda_1)^2\rangle &=& \left((\delta \Omega^2)^2 +  (2\delta K_{12})^2\right) \times \frac{3}{2 N} \\ 
    \langle(\delta \lambda_c)^2\rangle &=& \left((\delta \Omega^2)^2 +  (2\delta K_{12})^2\right) \times \frac{1}{2 N}, 
\end{eqnarray}
so the central eigenvalue is a factor $1/\sqrt{3}$ less sensitive to distortions (on average) than the most extreme. 
This is not a big difference per se, but the neighbouring mode is much closer in frequency for $m=1$ than for $m=c$ and  thus the possible misidentification problems on the former are much more severe. 

Also interesting is to study the mode distortion due to errors. 
The deviation of the eigenvectors of the imperfect system can be computed as a sum over the other unperturbed modes,  
\begin{equation}
\label{modedist}
    \delta {\bar e}_m 
= \sum_{m'\neq m} \frac{({\bar e}_{m'})^T {\delta \mathbb{M}} ({\bar e}_{m})}{ \lambda_{m}-  \lambda_{m'} }  {\bar e}_{m'} 
= \sum_{m'\neq m} \frac{({\bar e}_{m'})^T {\delta \mathbb{M}} ({\bar e}_{m})}{2k(\cos\beta_m - \cos\beta_{m'}) }  { \bar e}_{m'} ,
\end{equation}
where it is evident that modes closer in frequency will be much more severely distorted by neighbours than isolated ones. 
Since we have analytical formulae for the nearest-neighbours of the axion we can predict the hyperplane in the N-dimensional field-space in which the modes will tend to move more and thus estimate a reduction of the coupling to the axion mode,  
\begin{equation}
\label{deltaG}
    \delta {\cal G}_m = \frac{(1,...,1)}{\sqrt{N}}\cdot \delta \vec e_m . 
\end{equation}

As a general comment, note that for our optimised setups, where already one eigenvector is aligned with the axion vector $(1,1,..,1)$ the rest are orthogonal and thus $\delta {\cal G} =0$ at this linear order. Any distortion in ${\cal G}$ will enter at quadratic order, even if the mode distortions \eqref{modedist} are non-zero.  

Let us go back into our unoptimised setups to compare the purely inductive with the alternating filter. Again the main difference stems from the fact that in one case the axion mode is $m=1$ and in the other is $m=c\sim N/2+1$. As a further study of the deviation of the eigenvectors (\ref{modedist}) we can compute the expectation value of the variance of the numerator for both cases and we would find something relatively similar to the eigenvalue case, $O(1)$ differences. However, here the difference in denominator scales completely differently because as we discussed above, modes tend to group together around $m=1,N$ as $1/N^2$ while the distance between the central and neighbours scales as $1/N$. This is where the greatest advantage of our filters lie. 

As a technical aside, note that the most relevant distortion of the mode $m=1$ will happen in the direction of $m=2$, while for the mode $m=c$ the largest distortion happens in the plane spanned by the $c-1, c+1$ modes. 
However, interestingly neither of these three modes ($m=2,c-1,c+1$) couple to the axion direction and thus the most important contribution to the distortion of the geometric factor ${\cal G}_1$ will be due to $m=3$ and to ${\cal G}_c$ due to $c-2,c+2$.

Finally, note that similar procedures can be used to design new systems for which other eigenfrequencies of $\mathbb{M}_S$ are the ones coupling the most to the axion.   
However, for us it is very clear that the central mode is the most advantageous one. 


\subsection{Application to dielectric haloscopes}

Besides the application to the design of our RADES filters for axion research, the same principles should be applicable to other ideas existing in the literature. One example would be dielectric haloscopes, such as MADMAX~\cite{Brun:2019lyf} and LAMPOST~\cite{Baryakhtar:2018doz}. Dielectric haloscopes consist of a sequence of plane-parallel dielectric disks, potentially forming resonant cavities in between~\cite{TheMADMAXWorkingGroup:2016hpc}.

In particular, it was shown in reference~\cite{Millar:2016cjp} that in highly resonant configurations a dielectric haloscope should behave like a series of coupled cavities, and similar calculations should hold for both. To replicate the prototypical example in this paper (see section \ref{sec:alternating_experiment}), we require seven dielectric disks to form into six cavities. The only degree of freedom we can play with is the disk thicknesses, however it seems likely that the couplings caused by inductive and capacitive irises can be captured by alternating positive and negative phase changes of the waves propagating across the disks. Thus we will compare all disks with a phase thickness $\delta \sim \pi/2$ to alternating 
$\delta \sim \pi/2$ and $\delta \sim 3\pi/2$. As a test case, we choose a relatively low refractive index $n = 5$, and plot the possible variations: all distances $\delta \sim \pi/2$, all $\delta \sim 3\pi/2$, alternating starting with $\delta \sim \pi/2$ and alternating starting $\delta \sim 3\pi/2$. We have defined $\omega_0$ such that for disks at distance $d$ apart, $\omega_0=\pi/d$. 

\begin{figure}[!htb!]
\centering
  \includegraphics[width=.49\linewidth]{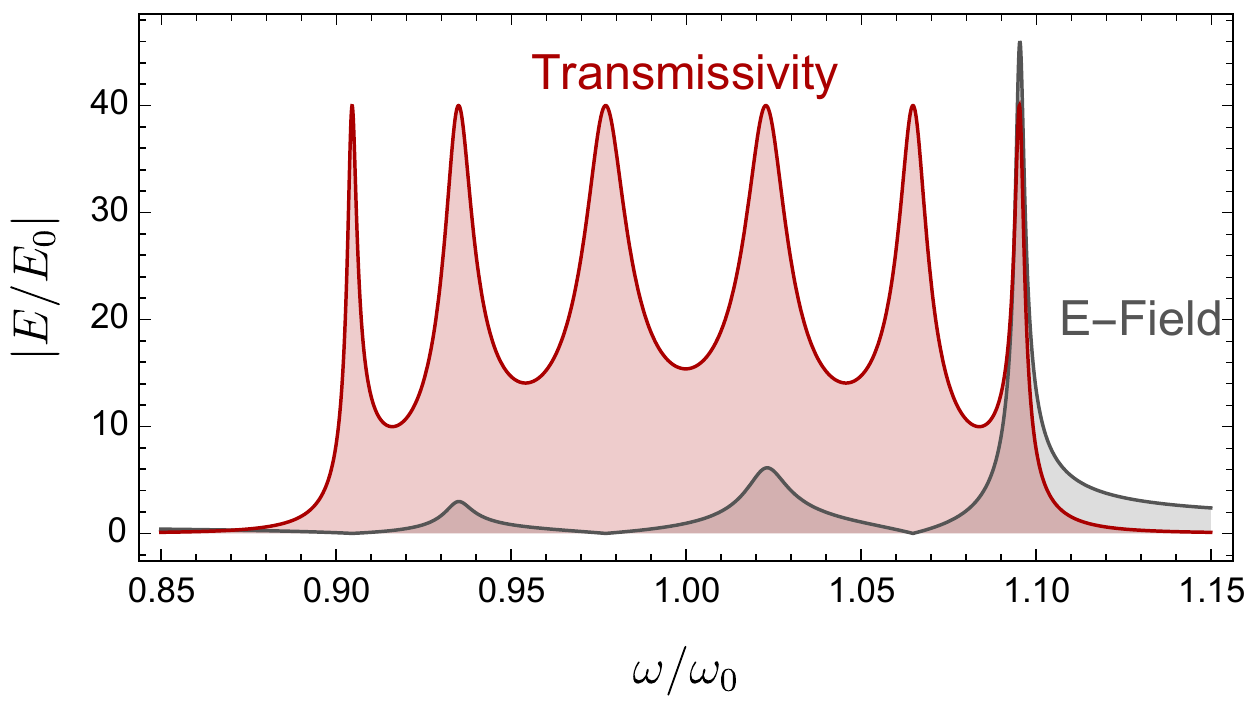}
  \includegraphics[width=.49\linewidth]{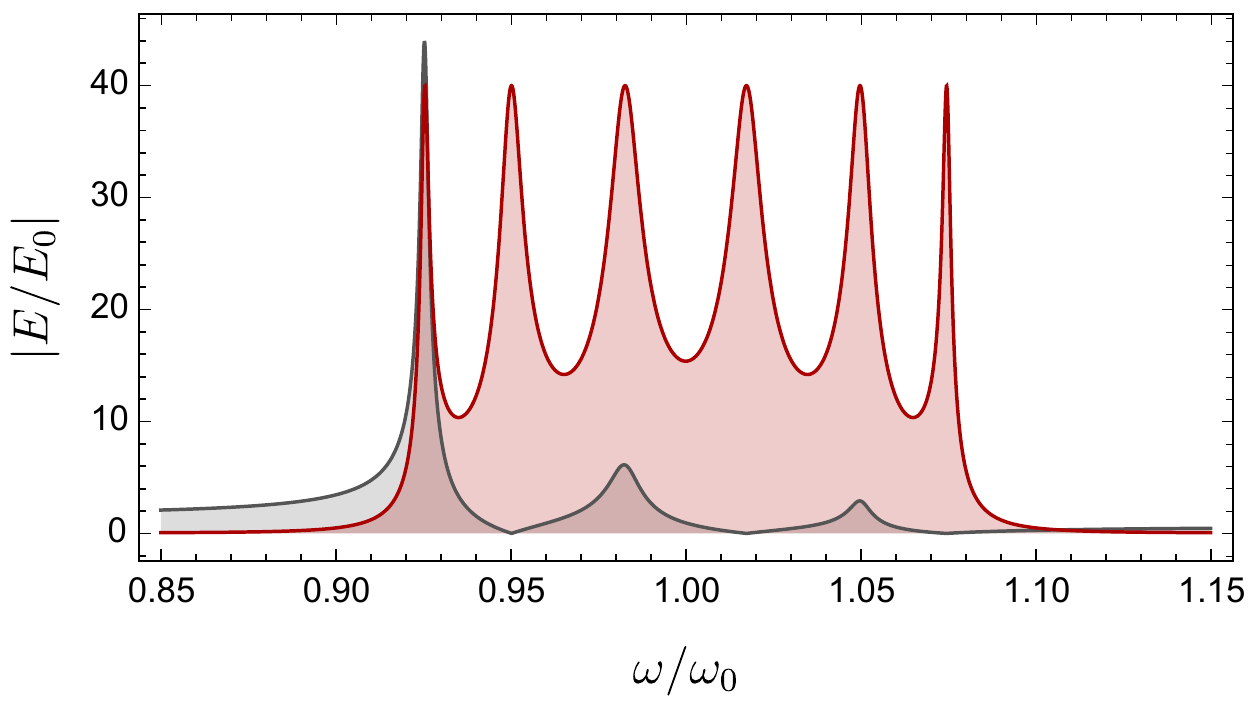}
    \includegraphics[width=.49\linewidth]{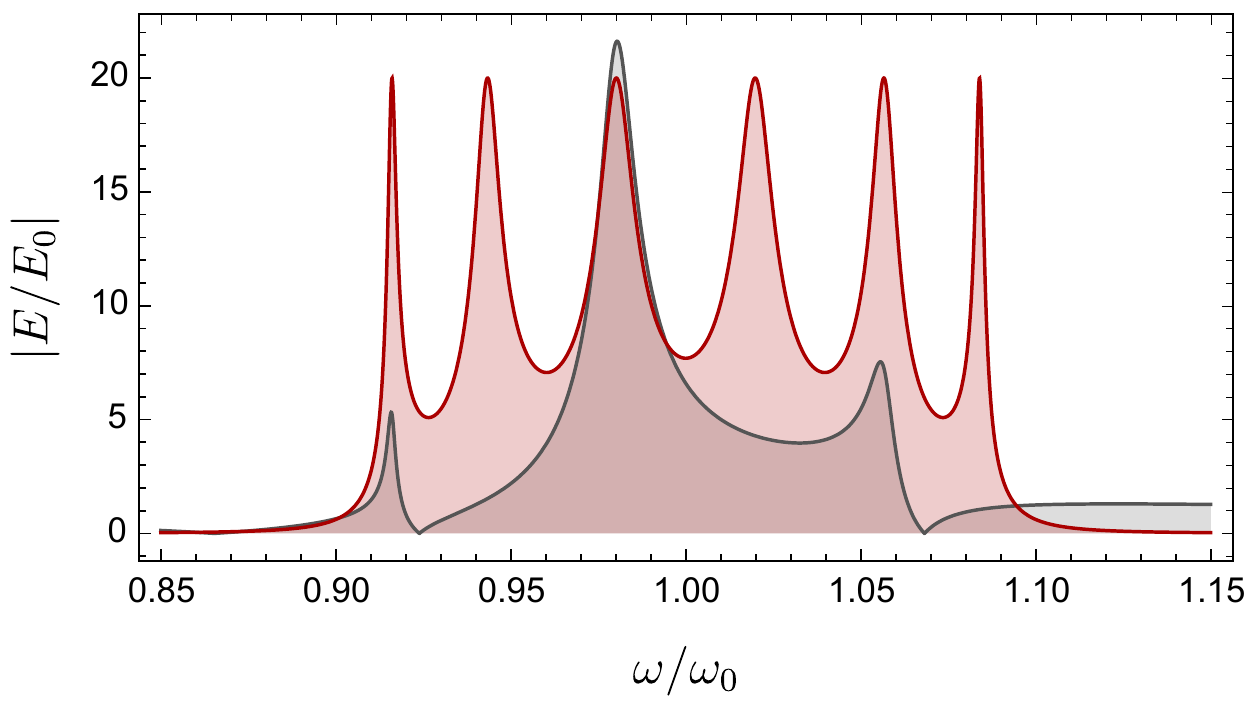}
      \includegraphics[width=.49\linewidth]{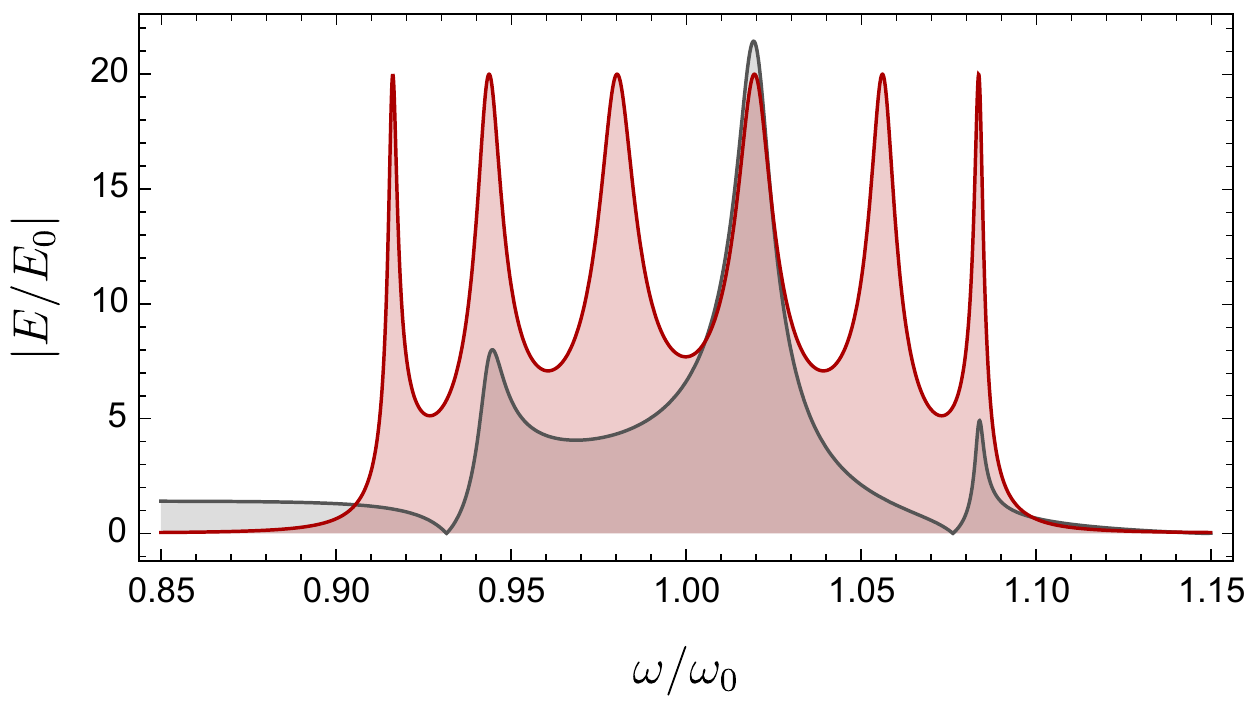} 
  \vspace{-.1cm}
\caption{Normalised output E-field (gray) and scaled transmissivity (red) for a dielectric haloscope consisting of seven dielectric disks. The refractive index is chosen to be $n = 5$, with each disk placed equidistantly $d$ apart, with $\omega_0=\pi/d$. The top left corresponds to all disks having phase thicknesses $\delta \sim \pi/2$. The top right has all $\delta \sim 3\pi/2$. The bottom left has alternating disks starting with $\delta \sim \pi/2$ and the bottom right has alternating disks starting $\delta \sim 3\pi/2$. The transmissivity of each setup is scaled up to match the E-field (by a factor 40 in the top panels, 20 in the bottom).\label{fig:n=5} 
}
\end{figure}

In Fig.~\ref{fig:n=5} we show the amplitude of the E-field of the outgoing wave normalised to $E_0=g_{A\gamma}B_eA_0\omega$, and transmissivity $T$ vs frequency for each case. Note that in the language of dielectric haloscopes the normalised outgoing E-field, $E/E_0$, is often referred to as the ``boost factor", and is calculated using the transfer matrix formalism of reference~\cite{TheMADMAXWorkingGroup:2016hpc}. In the limit that $n\to \infty$ the transfer matrix formalism should correspond to our linear matrix description, as shown by the similarity in results. The bottom panels, corresponding to the alternating cases, show that the axion couples most strongly to the third and fourth modes, as predicted by Eqn.~\eqref{eqn:altmodes}. We shall focus on the bottom right panel, in anticipation of the cavity studied in the next section. 
We can study the behaviour more carefully by looking at the mode structure that gives the bottom right panel. In general, dielectric haloscopes do not have cavity modes per se. Instead the coupling of dielectric haloscopes to axions can be found by solving for the reflection of a wave onto the system in the absence of the axion~\cite{Millar:2016cjp}. Such a procedure is equivalent to finding the free photon wavefunctions of the system, the so called ``Garibian wavefunctions"~\cite{Ioannisian:2017srr}. Thus to replicate the cavity mode studies we must look at reflectivity maps. Note that as the dielectric haloscope becomes more and more resonant, the reflectivity simply maps out the modes of the system~\cite{Millar:2016cjp}.

\begin{figure}[!htb!]
	\centering
	  \includegraphics[width=.45\linewidth]{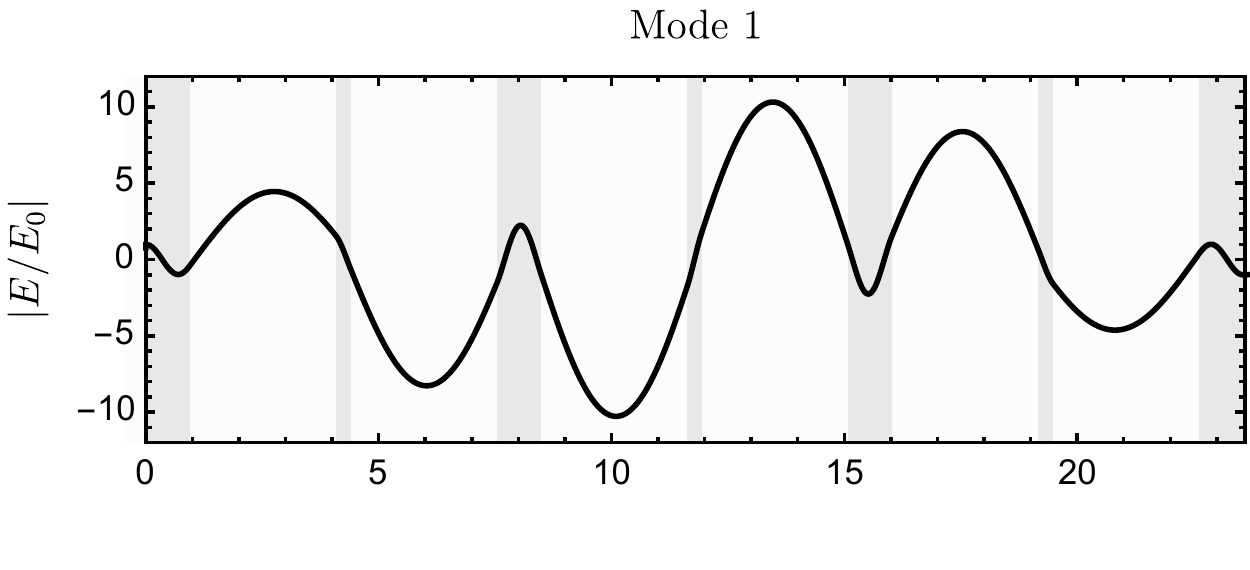}
	  \includegraphics[width=.45\linewidth]{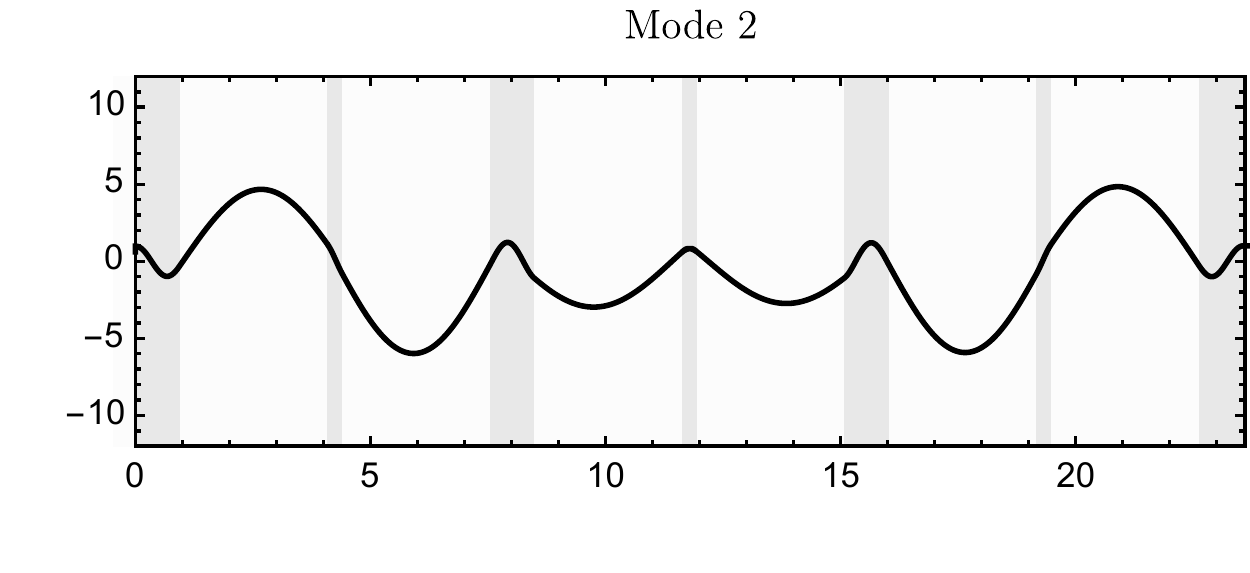} 
	    \includegraphics[width=.45\linewidth]{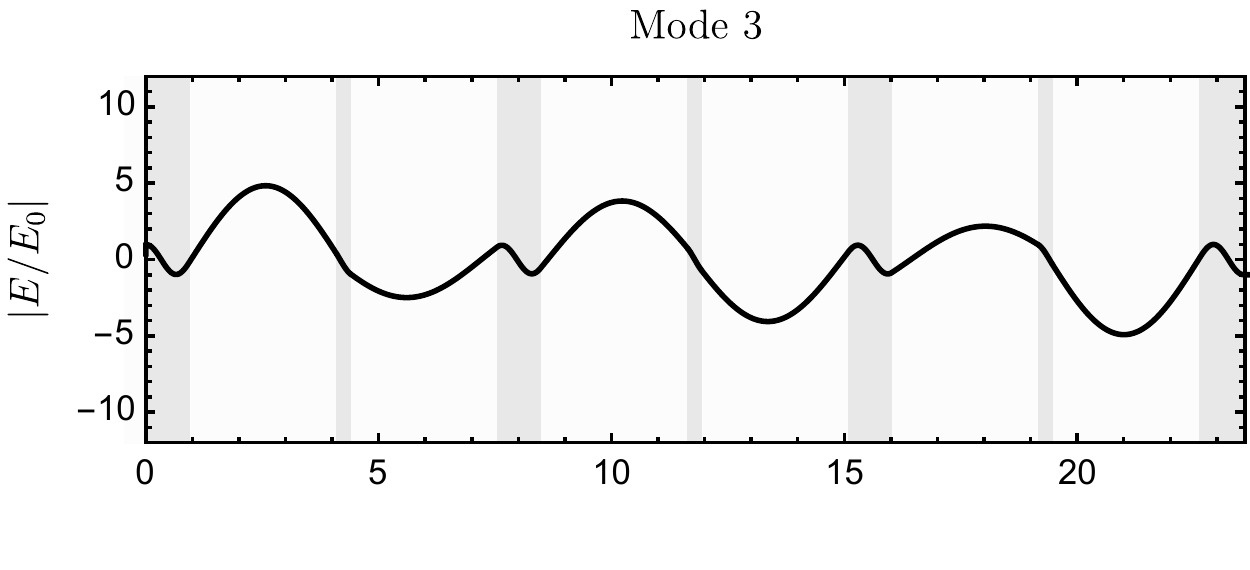}
	      \includegraphics[width=.45\linewidth]{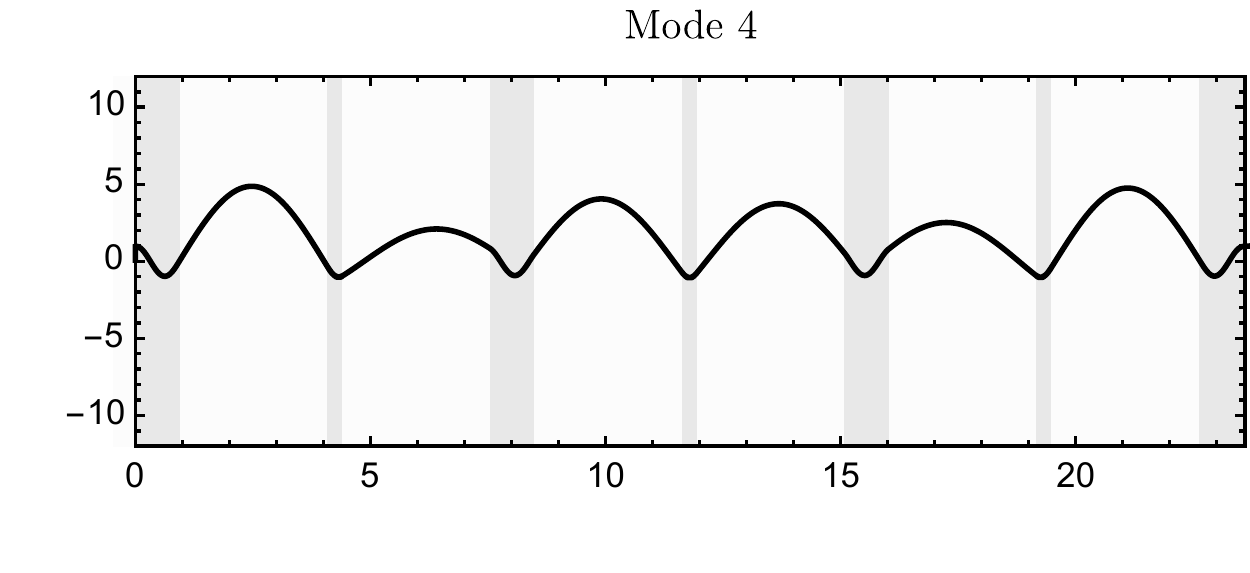} 
	      \includegraphics[width=.45\linewidth]{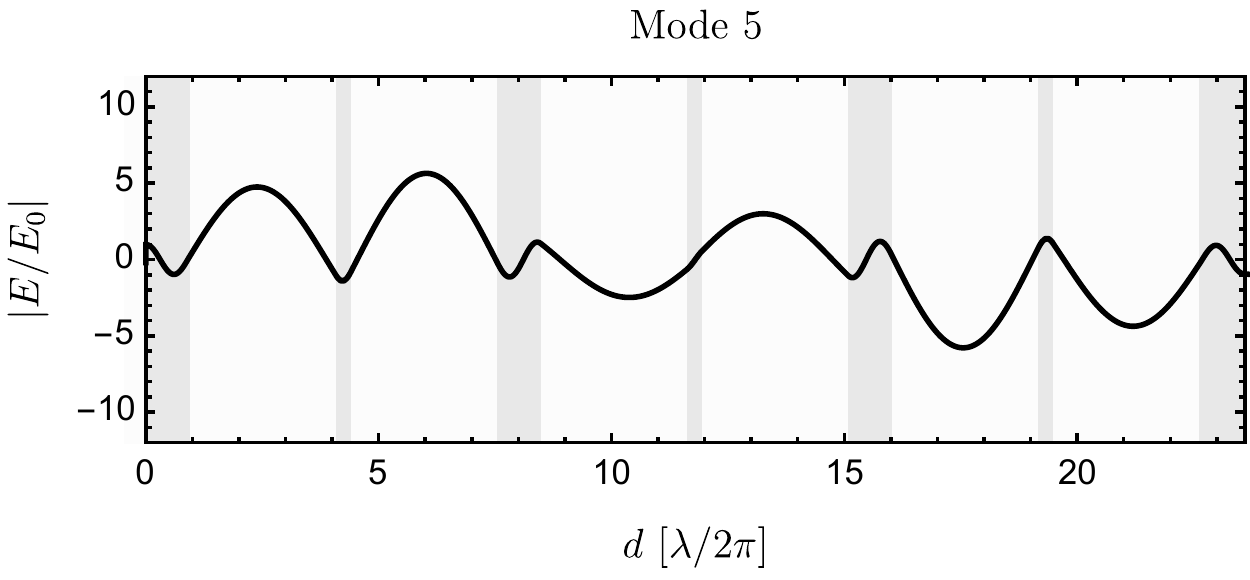}  
	      \includegraphics[width=.45\linewidth]{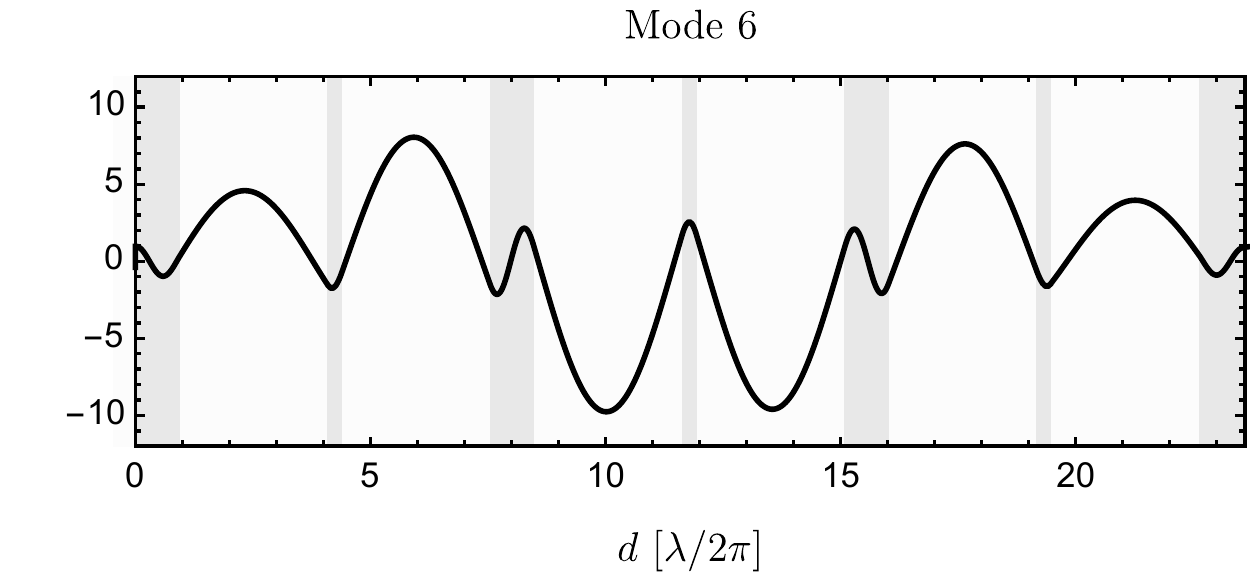} 
	  \vspace{-.1cm}
	\caption{E-field distribution excited by a reflected wave as a function of distance $d$ from the left hand side of the dielectric haloscope in the bottom left panel of Fig.~\ref{fig:n=5}. When integrated this E-field gives the power generated by the axion field in the system. As labeled, the panels refer to the frequencies of the six modes shown in the transmissivity of Fig.~\ref{fig:n=5}.
In each panel the locations of the dielectric disks are indicated respectively by the light-gray vertical bars.
\label{fig:modes}
	 }
	\end{figure}

As shown in Fig.~\ref{fig:modes}, the mode structure has a remarkable similarity to the analytical description in \eqref{Wmodes0} and the modes of the optimised filter that we will see later in Fig.~\ref{all_modes_E}, especially considering that the system is only mildly resonant. Thus we can be confident that a dielectric haloscope with alternating disk thicknesses captures the essential behaviour of the RADES system of alternating couplings. As such, it will share its advantages, particularly a smaller sensitivity to manufacturing errors. Note that this is particularly critical in the MADMAX experiment because the distance between disks will be actively moved and chosen at small time intervals during data taking. Thus less sensitivity to mechanical tolerances translates into less sensitivity to positioning errors and thus the possibility of using larger boost factors~\cite{Millar:2016cjp}.
In general, the disks used in a dielectric haloscope like MADMAX will not be a simple phase thickness like $\pi/2$. Realistic setups also use highly optimised spacings between the disks, rather than equidistant disks. Thus it is not initially clear that the lower sensitivity to errors can be achieved in these setups. Because of this we now turn our attention to realistic setups, similar to those studied in \cite{TheMADMAXWorkingGroup:2016hpc}. Rather than disks which are simply $\pi/2$ and $3\pi/2$ thick, we will consider alternating disks chosen to reduce the sensitivity to error. We will consider the case of 20 dielectric disks with a mirror on one side, as will be used in the prototype for MADMAX~\cite{Brun:2019lyf}.

\begin{figure}[!htb!]
\centering
  \includegraphics[width=.49\linewidth]{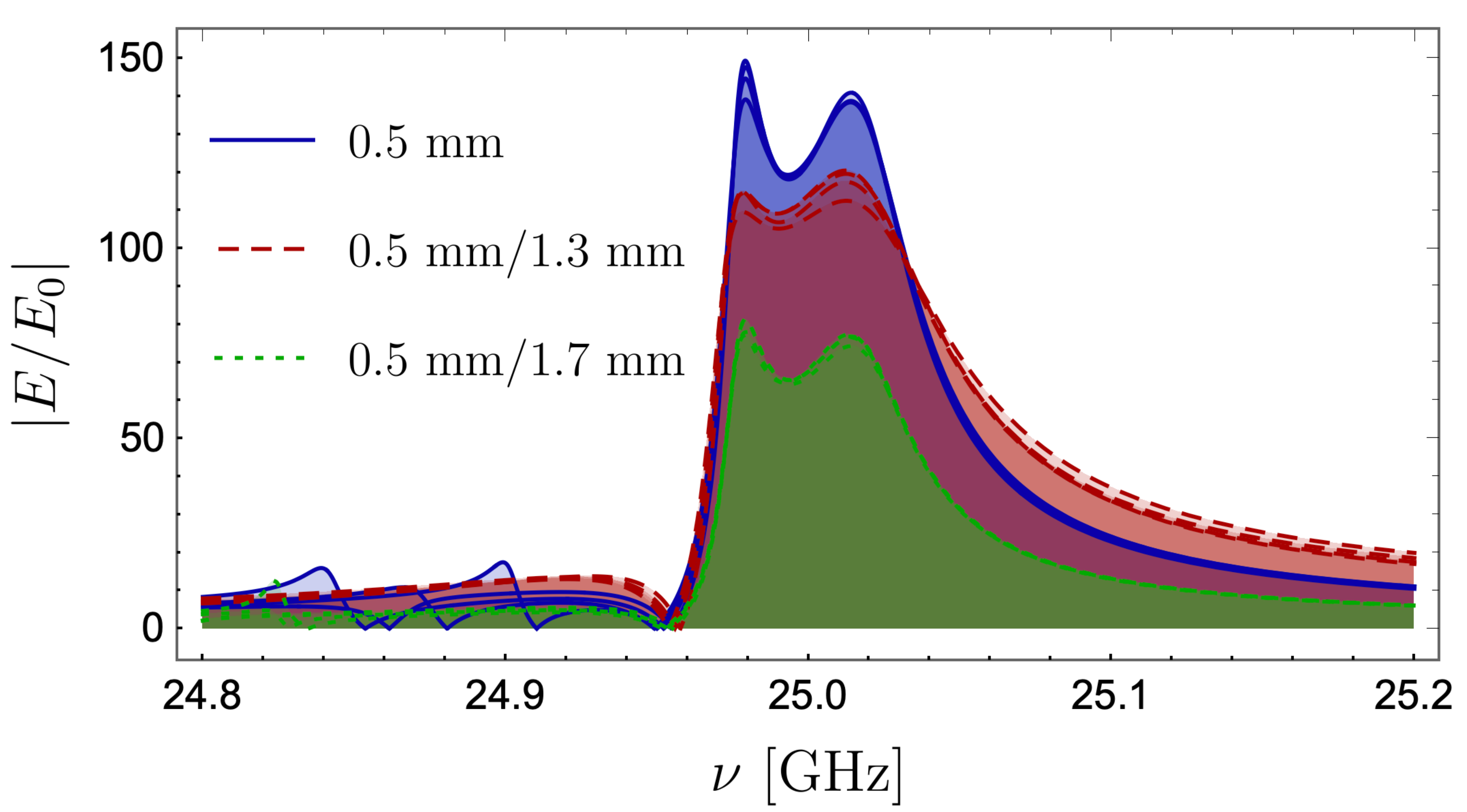}
  \includegraphics[width=.475\linewidth]{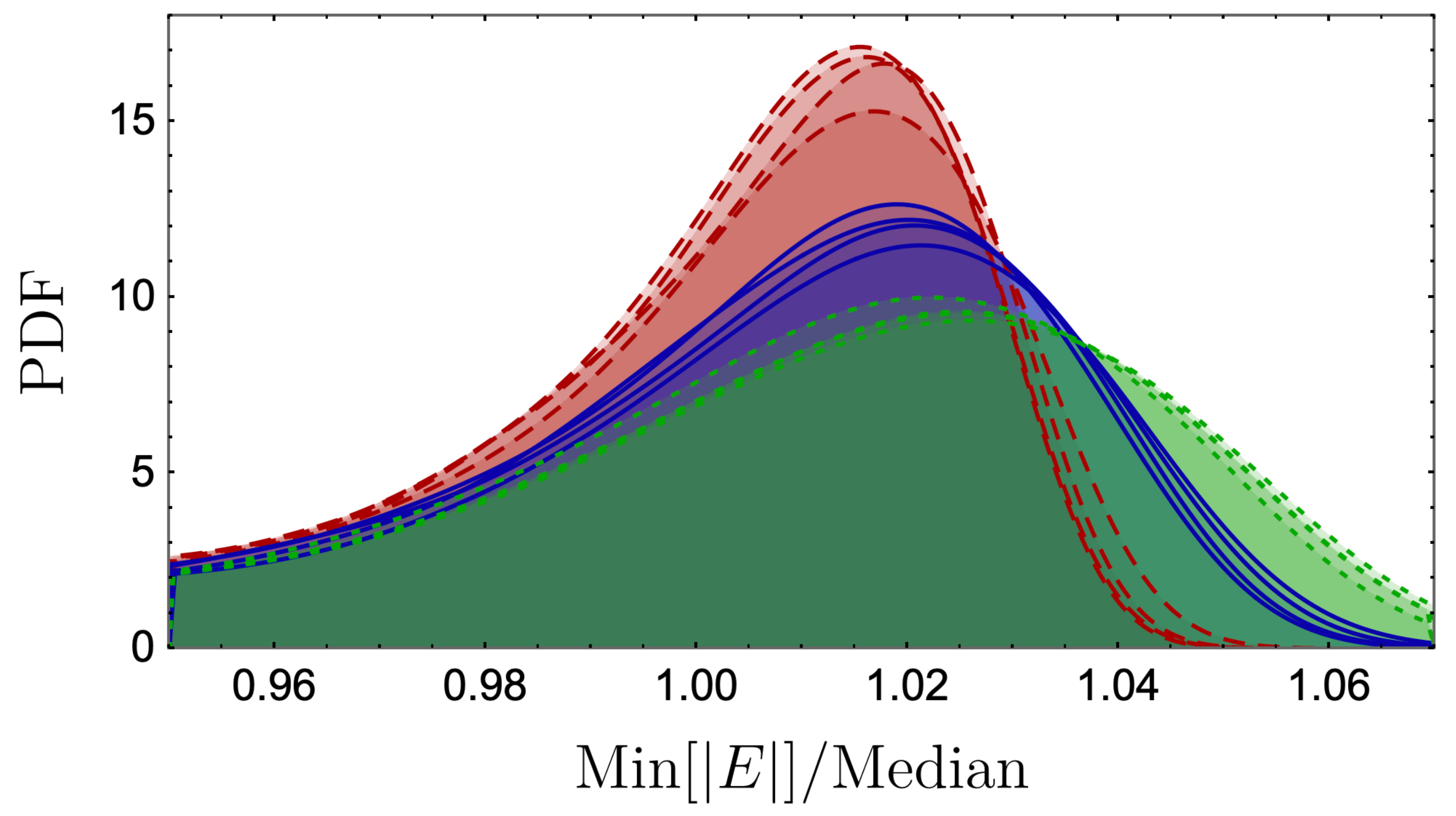}
  \vspace{-.1cm}
\caption{{\em Left:} Normalised outgoing E-field as a function of frequency $\nu$ for four dielectric haloscopes consisting of a mirror and 20 dielectric disks. The outgoing E-field has been optimised four seperate times for a 50 MHz bandwidth by adjusting the spacings between disks. The disk thicknesses are given by 0.5~mm for all disks (blue), alternating 0.5~mm with $\sim 1.3~$mm disks (red dashed) and alternating 1~mm and $\sim 1.7~$mm disks (green dotted). 
{\em Right:} Probability density as a function of minimum achieved E-field within the bandwidth for each of the three above dielectric haloscopes (same colouration). The PDF is estimated by perturbing the disks spacings used in the left panel by a top hat error function of width 5~$\mu$m 10,000 times. To account for the different E-fields, we normalise by the median value, showing qualitatively different behaviours between the different studied cases.    
}\label{fig:realistic} 
\end{figure}

In the left panel of Fig.~\ref{fig:realistic} we show the normalised E-field as a function of frequency for three dielectric haloscopes. As the space of disk spacings is high (20) dimensional, optimisers generally find local, rather than global, maxima. To disentangle the effect of carefully chosen disk thicknesses, we show the cases of 0.5 mm disks, 0.5~mm alternating with $\sim1.3$~mm and 0.5 mm alternating with $\sim1.7$~mm. The spacings between disks are optimised to maximise the power output in a 50~MHz bandwidth, centred on 25~GHz. Each case is optimised four times from scratch to indicate generic behaviour. Due to the differing disk thicknesses, we observe a range of ``boost factors''. To test the sensitivity to disk mispositioning errors, we perturbed the optimised positions 10,000 times by a top hat error function of 5~$\mu$m. In the right panel of Fig.~\ref{fig:realistic} we estimate the probability density function (PDF) for the minimum achieved E-field using these perturbations, normalising by the median value.  Depending on the chosen thicknesses different error sensitivity is achieved, unpredicted by the achieved boost factor. In particular, the case with 0.5~mm and $\sim1.7$~mm disks actually is more sensitive to errors despite the lowest boost factor by far. On the other hand, while the case with 0.5~mm and 1.3~mm disks has a similar boost factor to homogeneous disks, it is approximately 30\% less sensitive to errors. Thus we see even in realistic setups alternating disk thickness can lead to significantly better tolerance to mechanical faults. Note that we did not choose a special bandwidth or initial disk thickness for these results; once a pair of alternating thicknesses that reduce error sensitivity are found, they appear to work for any bandwidth. Further, the existence of such alternating thicknesses is generic, though due to the complicated nature of the setups there does not seem to be a simple prediction for which thicknesses are optimal.

Just like RADES, dielectric haloscopes like MADMAX can benefit from optimising the couplings between different parts of the system. Given the high number of degrees of freedom and the relative transparency of the materials in the presently studied case, the analogy with the RADES alternating cavity design does not hold perfectly. One potential difficultly is that the phase relationships between different disks only hold for a given frequency. Because of this it may be difficult to reduce the sensitivity to mechanical errors for all frequencies. However, these results do motivate the designers of dielectric haloscopes to optimise not only for high power, but also for robustness; it is possible that, with optimised disk thicknesses, dielectric haloscopes can be made robust against disk mispositioning for a wide range of frequencies. Note that existing designs for dielectric haloscopes generally assume at least two sets of disks~\cite{Millar:2016cjp}, with the exact number of required sets not fully studied. We leave the detailed study of practical designs of dielectric haloscopes to future work.

\section{Design and performance of the cavity filter with alternating irises \label{sec:alternating_experiment}}

\subsection{Electromagnetic design}
\label{s:electro_design}
In the following we describe the design procedure of a prototype formed by the cascade connection of $N=6$ cavities using alternating inductive and capacitive irises.

The first design specification is the resonating frequency $f$ which sets the mass of the axion searched for. We chose to work at a frequency of $f\sim\unit[8.5]{GHz}$.

Imposing a $(1,1,1,1,1,1)$ eigenvector, which represents the desired mode
distribution, and using the formulation discussed in section \ref{sec:alternating_theory}, we can select the resonating frequencies in each sub-cavity. By assuming that the fundamental $TE_{101}$ mode is resonating in each elemental resonator, the dimension of the cavities can be easily calculated with the well known expression,
\begin{equation}
\label{length_cavity}
f \, = \, \frac{c}{2} \, \sqrt{\frac{1}{a^2} + \frac{1}{l^2}},
\end{equation}
where $c$ is the speed of light in vacuum, $l$ the length and $a$ the width of the rectangular waveguide sub-cavities. In this application we have used a standard WR-90 rectangular waveguide, where $a=\unit[22.86]{mm}$ and $b$, the height of the elemental resonators, is $b=\unit[10.16]{mm}$.
It should be pointed out that for the alternating (inductive and capacitive irises) design, the solution imposes equal lengths for the internal cavities, and a slightly different length for both sub-cavities at the ends. This length variation compensates for the fact that the internal cavities are connected to two irises while the ending cavities to a single one
(see \cite{Melcon:2018dba} for a detailed derivation).

\begin{figure}[h!]
\begin{center}
\includegraphics[width=0.45 \textwidth]{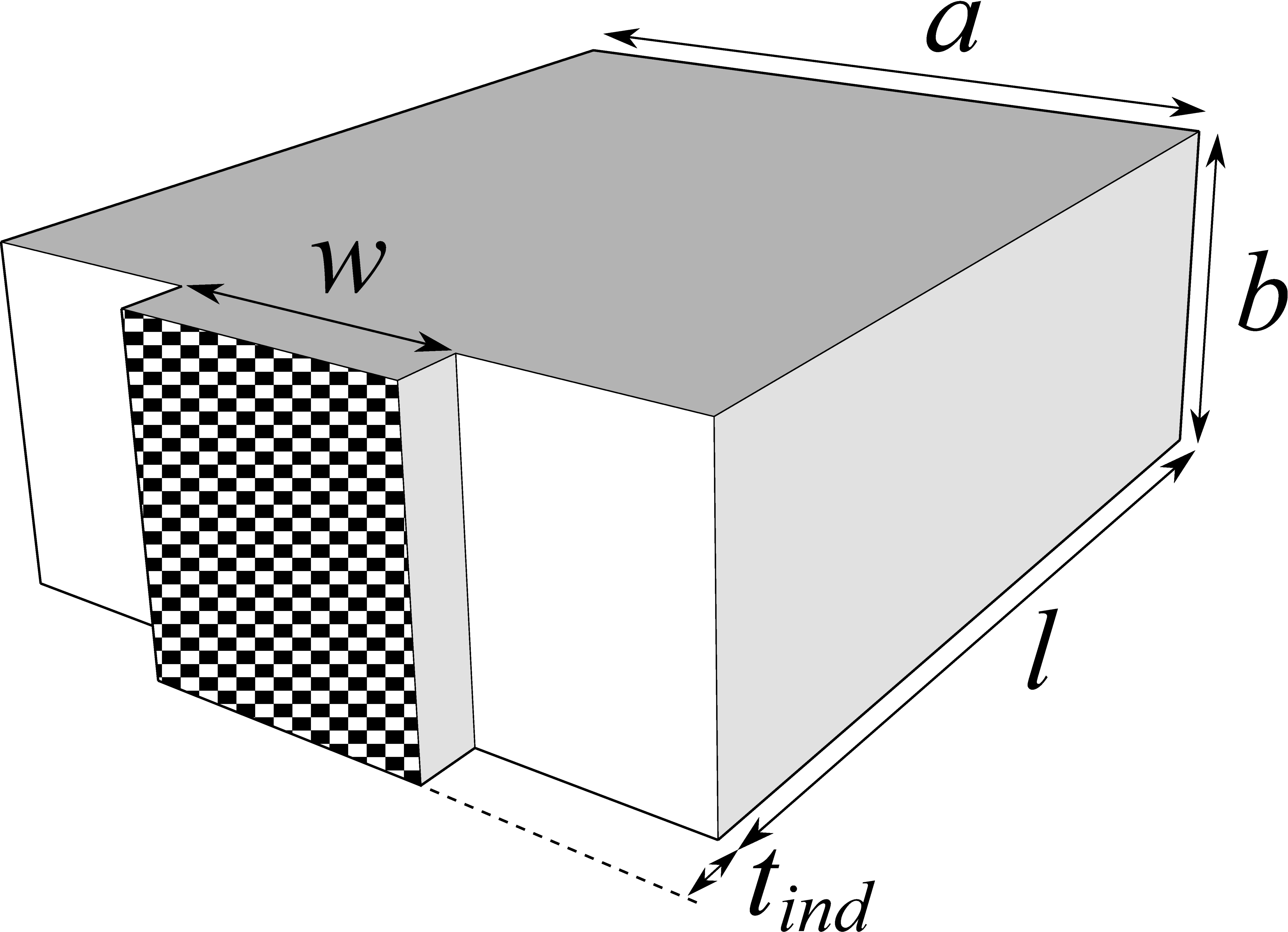}
\includegraphics[width=0.6 \textwidth]{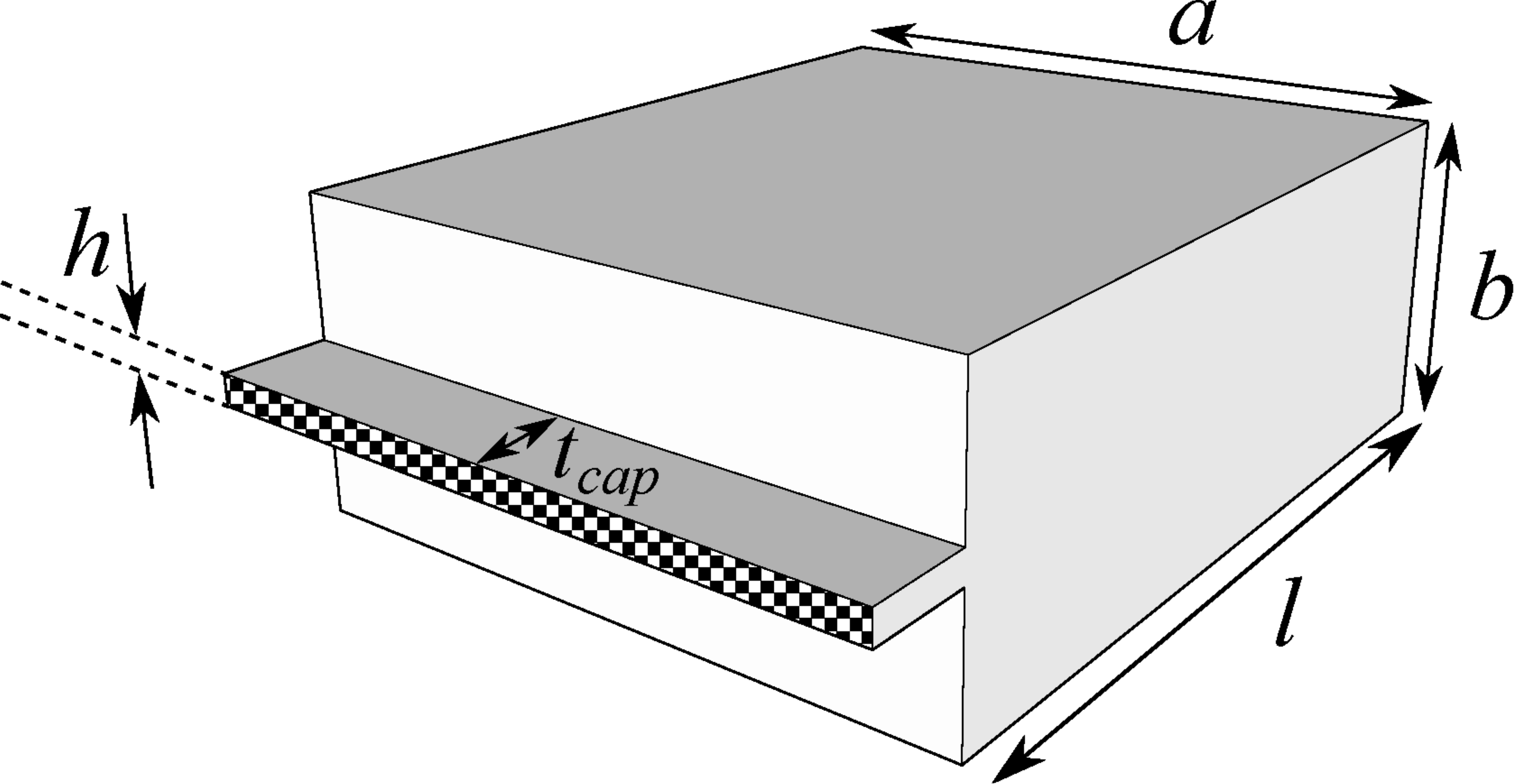}
\caption{\label{inductive_capacitive} Layout of the inductive (top) and capacitive (bottom) irises. The hatched zones indicate the connection planes between the irises and the cavity.}
\end{center}
\end{figure}

Figure~\ref{inductive_capacitive} depicts the design of both inductive and capacitive irises. They are characterised by the width and the length of the iris, $w$ and $t_{ind}$, for the former and the height $h$ and the length $t_{cap}$ for the latter.

The minimal dimensions of $t_{ind}$ and $t_{cap}$ are dictated by mechanical and manufacturing constraints. We have chosen to use $t_{ind} = \unit[2.0]{mm}$ and $t_{cap} = \unit[3.5]{mm}$.\\
The choice of coupling factors will impact the remaining dimensions of the irises. In order to choose appropriate coupling factors $k$ taking into account geometrical constraints, we calculated the $k$ factors following the procedure described in section 14.2 of \cite{cameron}, which is based on an impedance (admitance) inverter equivalent circuit of two coupled resonators in the case of the inductive (capacitive) iris. In both cases the resonator is represented as the simple connection of a capacitive and an inductive lumped element. By connecting two identical cavities through an inductive or capacitive iris, and imposing `Perfect Electric Conductor Wall' (PEC) and `Perfect
Magnetic Conductor Wall' (PMC) boundary conditions in the central plane of the iris, which is perpendicular to the energy transmission direction, one can compute the even (PMC) $f_e$ and odd (PEC) $f_o$ resonant frequencies. The relationship between the coupling factor $k$
and these two frequencies is \cite{cameron}
\begin{equation}
|k| =  \frac{|f_e^2-f_o^2|}{f_e^2+f_o^2}.
\label{eq:k_eq}
\end{equation}
We note that in the case of inductive irises, $f_e^2 < f_o^2$, the coupling coefficient is thus negative and we express it as $k_2=-|k|$. For capacitive irises, $f_e^2 > f_o^2$ and we thus note it as $k_1=|k|$.\\
 CST Microwave Studio \cite{cst} was used to carry out this electromagnetic analysis using the eigensolver simulation module. In the electromagnetic simulation the cavity is considered to be made entirely of copper, although in fact the cavity is made of stainless steel for mechanical resistance to magnet quench but with a copper layer of $\unit[30]{\mu m}$ to provide high electrical conductivity.
In the simulation we can use copper only because
the skin effect ($\delta_{300K} = \unit[0.716]{\mu m}$) is two orders of magnitude smaller than the thickness of the copper layer ($\unit[30][\mu m]$) used. Moreover,the penetration of the fields at $\unit[2]{K}$ is limited by the anomalous regime, which sets in when the skin depth is equivalent to the electron mean free path at about $\unit[0.27]{\mu m}$, which is also much smaller than the copper thickness, see e.g. \cite{finger08}.

\begin{figure}[h!]
\includegraphics[width=.5\textwidth]{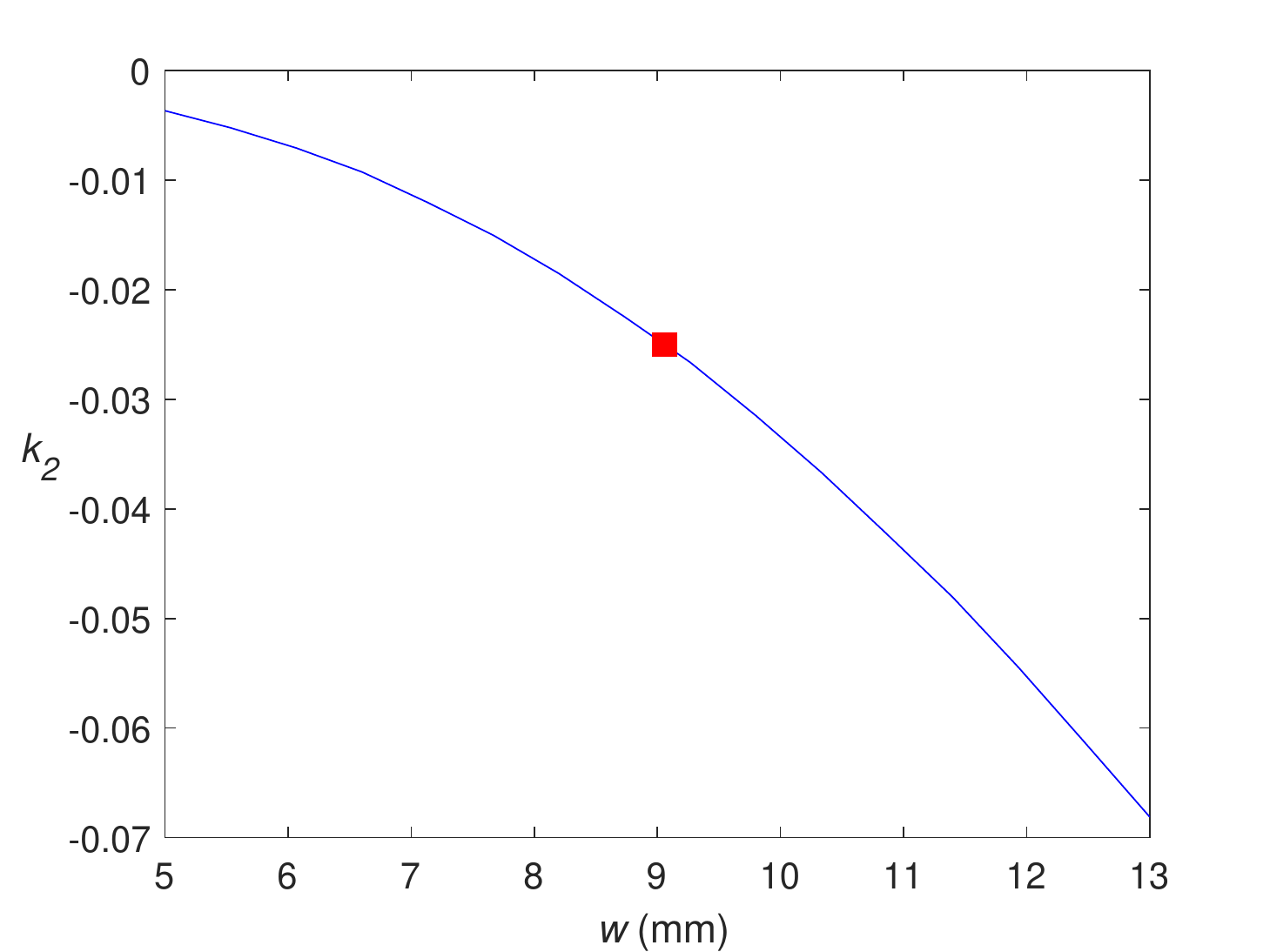}
\includegraphics[width=.5\textwidth]{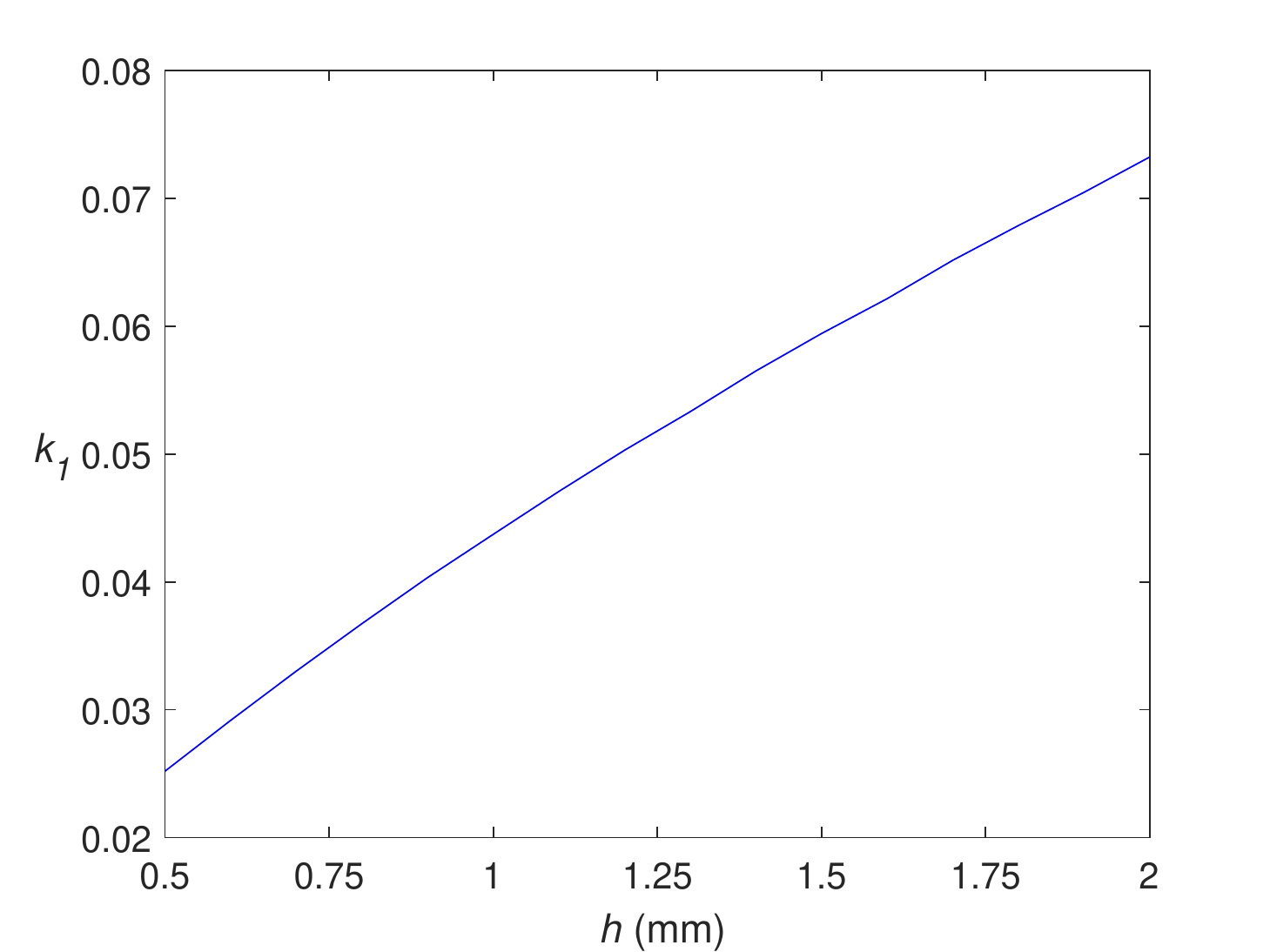}
	\caption{\label{k_inductive_capacitive} 
Coupling coefficients for the inductive (left) and capacitive (right) irises using $t_{ind} = \unit[2.0]{mm}$ and $t_{cap} = \unit[3.5]{mm}$, respectively. The square point in the left plot indicates the chosen value for the normalized coupling $k$ of the inductive iris.}
\end{figure}

At room temperature the electrical conductivity of copper is assumed to be $\sigma_{RT} = \unit[5.8\times10^7]{S/m}$. However, since the final experiments will be conducted at cryogenic temperatures we performed the simulations for a cavity  at \unit[2]{K}. At lower temperatures, the conductivity goes up but is limited by the Relative Resistivity Ratio (RRR) which, for \unit[2]{K}, is expected to be  between $30$ and $200$. We have chosen for our simulations $\sigma_{\unit[2]{K}} = 2.008\times\unit[10^9]{S/m}$ which assumes RRR$\sim 30$.\\
Figure~\ref{k_inductive_capacitive} shows the  coupling coefficient as a function of $w$ and $h$ for the inductive and the capacitive irises, respectively.

\begin{figure}[h!]
\includegraphics[width=0.5 \textwidth]{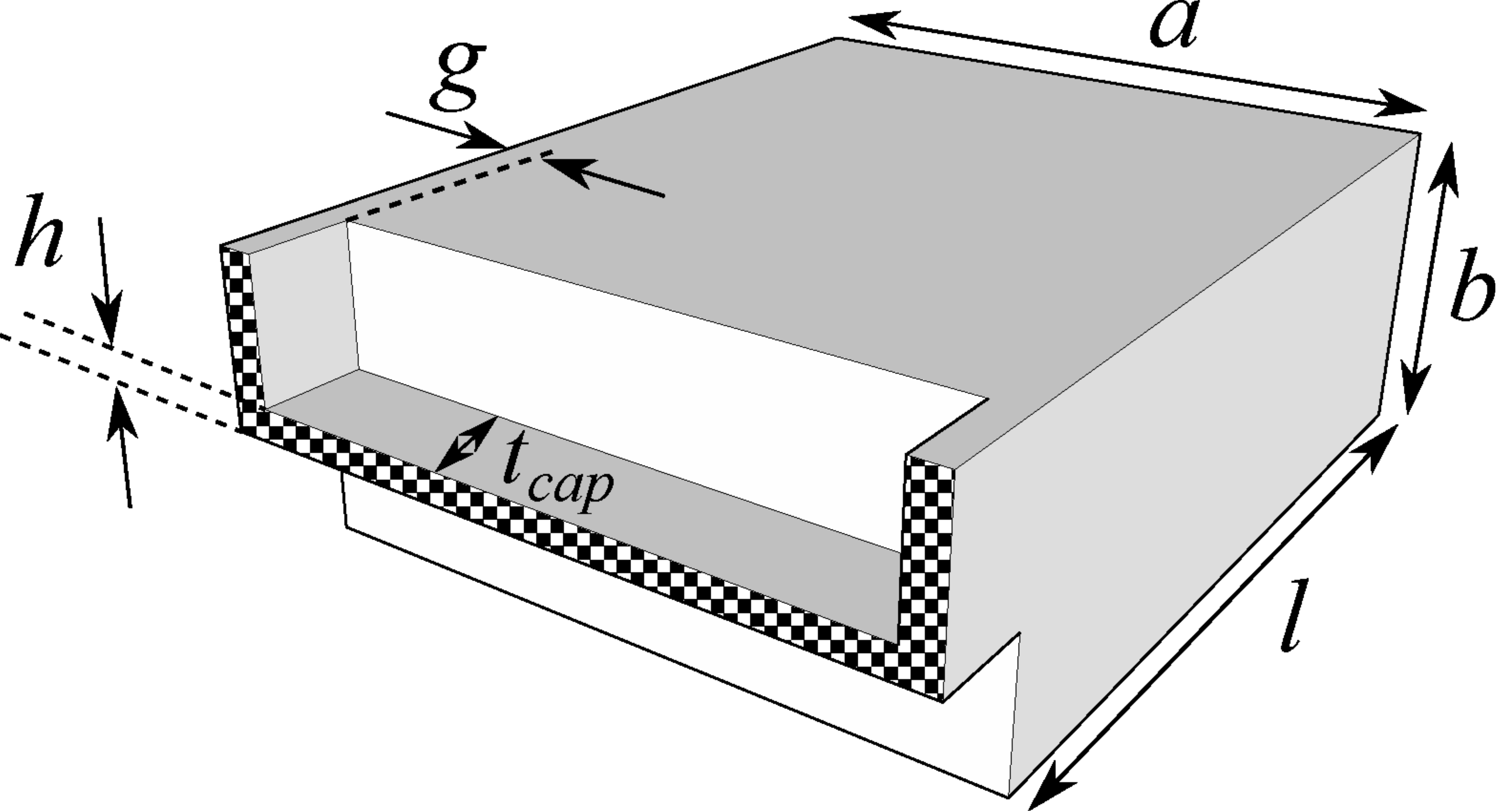}
\includegraphics[width=0.5 \textwidth]{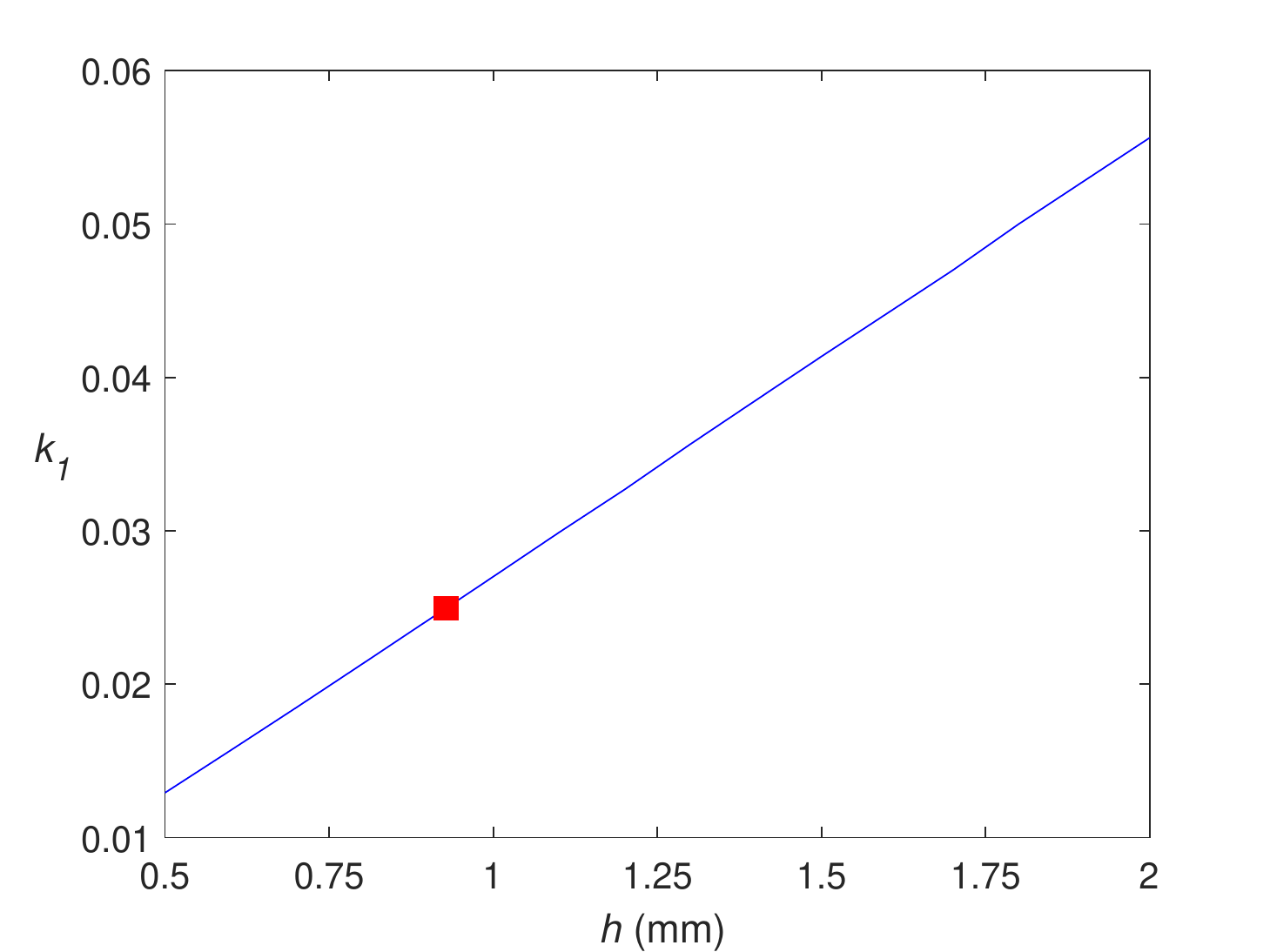}
	\caption{\label{k_cuasi-capacitive} 
Geometry (left) and coupling coefficient (right) as a function of $h$ of the quasi-capacitive iris using $t_{cap} = \unit[3.5]{mm}$ and $g =\unit[1.1]{mm}$. The hatched zone indicates the connection plane between the iris and the cavity. The square point indicates the chosen value for $k$.}
\end{figure}

Mechanical constraints for the production of the capacitive irises led to a modified geometry including protrusions of width \unit[1.1]{mm}, see the left plot of Figure~\ref{k_cuasi-capacitive}. Taking into account these protrusions lead to a modified relation between $k$ and $h$ as shown in the right plot of Figure~\ref{k_cuasi-capacitive}.\\
Based on our previous experience in the design of the fully-inductive RADES structure \cite{Melcon:2018dba}, we selected a value of $k$ similar to that case (which had been -0.0185), which was however slightly increased in order to ensure a reasonable value
for the height ($h$) of the capacitive irises. 

The final chosen value for
the coupling is indicated with square markers in Figure~\ref{k_inductive_capacitive} (left) and Figure~\ref{k_cuasi-capacitive} (right) and is $\pm |k| = \pm 0.0248$, resulting in $w = \unit[8.90]{mm}$ and $h = \unit[0.93]{mm}$.
 
The values of the four geometrical parameters:  length of the end-cap cavities ($l_{ext}$), length of the four internal cavities ($l_{int}$), width of the two inductive irises ($w$), and height of the three capacitive irises ($h$) were  varied in a final optimisation stage. The optimisation was performed with respect to the geometric form factor of the central mode ${\cal{G}}_4^2$ using the optimisation module available on CST Microwave Studio.

We note that the
optimisation process converged rapidly, since the initial design
provided by the described procedure was very close to
the desired specifications.\\

Two standard $50 \, \Omega$ SMA coaxial connectors were inserted in the first and in the last cavities. The first one, denoted as port $1$, will be used for the extraction of the axion signal; it has been designed for critical coupling regime at cryogenic temperatures. The second coaxial connector (port $2$) has been included for calibration purposes; the length of its internal pin has been reduced to its minimum physical value (undercoupled regime) in cryogenic operation mode. 

\begin{figure}[h!]
\includegraphics[width=\textwidth]{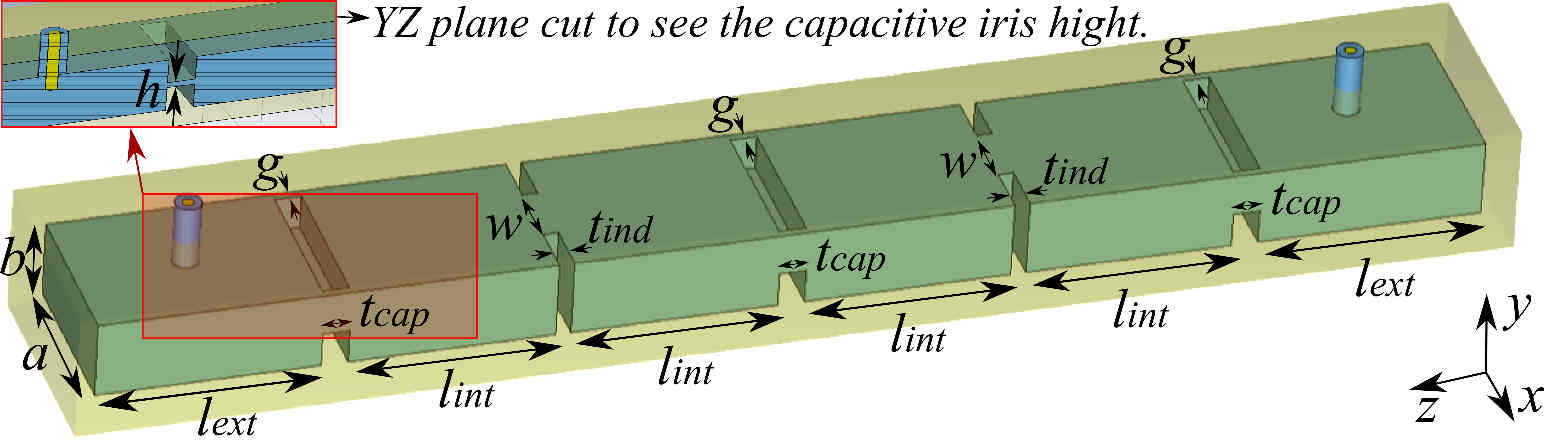}
\begin{center}
	\caption{\label{scheme_complete_filter} 
 Scheme of the complete filter including the coaxial probes. The yellow area represents the thickness of the structure's body made of copper-coated stainless steel.}
 \end{center}
\end{figure}

In Figure~\ref{scheme_complete_filter} we show the final design of the complete filter including the coaxial connectors. The geometrical dimensions at cryogenic and room temperatures, taking into account thermal expansion, are given in Table~\ref{physical_dimensions}.

\begin{table}[]

	\begin{tabular}{|c|c|c|c|}
		\hline
		Parameter  & $T = 2$ K & $T = 300$ K & $T = 300$ K \\
		 & (design) & (thermally expanded from \unit[2]{K}) & (manufactured) \\ \hline
		cavity width ($a$) & 22.860  & 22.929  & 22.860 \\ \hline
		cavity height  ($b$) & 10.160  &  10.191 & 10.272 \\ \hline
		length external cavities ($l_{ext}$) &  29.400 & 29.489 & 29.567 \\ \hline
		length internal cavities  ($l_{int}$) & 27.000 & 27.081 & 27.167 \\ \hline
		inductive iris width ($w$) & 8.900 & 8.987 & 8.986 \\ \hline
		inductive iris thickness  ($t_{ind}$) & 2.000 & 2.006 & 1.939 \\ \hline
		capacitive iris height  ($h$) & 1.000 &  1.063 & 1.019 \\ \hline
		capacitive iris thickness	($t_{cap}$)	& 3.500 & 3.511	& 3.436 \\ \hline
		capacitive iris	gap   ($g$) & 1.100  & 1.104 & 1.192 \\ \hline
	\end{tabular}
		\centering
	\caption{\label{physical_dimensions} Design physical dimensions (in mm) of the filter at \unit[2]{K} (first column), room temperature dimensions calculated taking into account thermal expansion (second column). The third column indicates the physical dimensions at room temperature of the manufactured cavity (see section \ref{s:Fabrication}) measured with an uncertainy of $\unit[3]{\mu m}$.} 
\end{table}

\begin{figure}[h!]
\includegraphics[width=0.9\textwidth]{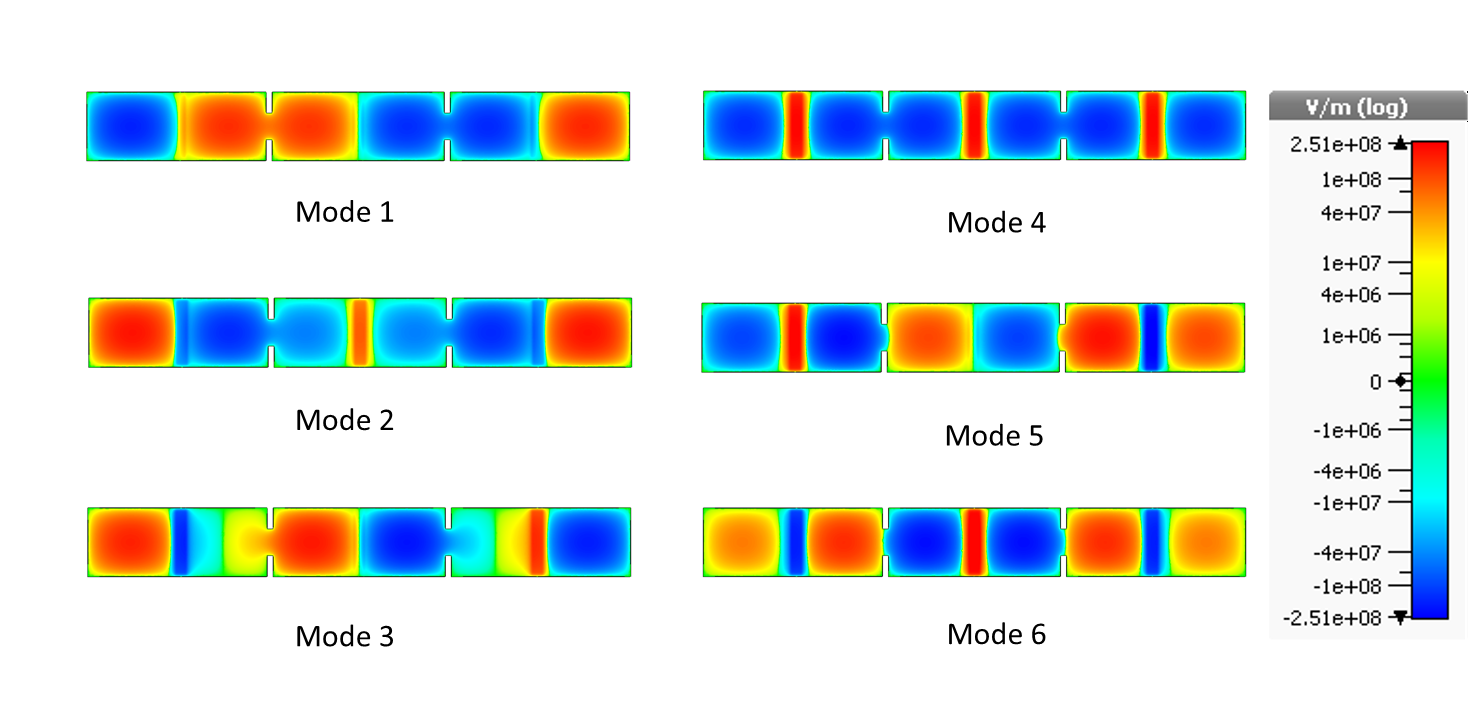}
	\caption{\label{all_modes_E} 
Electric field distribution of the six modes of the filter.
As visible, mode 4 resonates in phase inside the cavities, and thus couples maximally to the axion field.}
\end{figure}

The electric field patterns of the six cavity modes and their corresponding electromagnetic properties are depicted in Figure~\ref{all_modes_E} and Table~\ref{frequency_cavity_modes}, respectively. We see that the fourth mode exhibits a good alignment of the electric fields in all sub-cavities, while the field is inversely aligned only in the three small capacitive irises. Thus, mode $4$ constitutes the suitable mode for axion dark matter search, as it can also be seen numerically by computing the geometric form factor, shown in the third column of Table~\ref{frequency_cavity_modes} (following the procedure discussed in \cite{Melcon:2018dba}). 

\begin{table}[]
	\begin{tabular}{|c|c|c|c|c|}
		\hline
		Mode ($m$) & Resonant frequency (GHz) & ${\cal{G}}_m^2$ & $Q_0$ (at \unit[2]{K}) & $Q_0$ (at \unit[300]{K}) \\ \hline
		1 & 8.298 & 1.820 $\times 10^{-9}$ & 42878 & 7300 \\ \hline
		2 & 8.319 & 1.257 $\times 10^{-3}$ & 43758 & 7448 \\ \hline
		3 & 8.395 & 1.007 $\times 10^{-8}$ & 43626 & 7426 \\ \hline
		4 & 8.508 & 0.534 & 40657 & 6957 \\ \hline
		5 & 8.613 & 1.511  $\times  10^{-8}$ & 44168 & 7563 \\ \hline
		6 & 8.698 & 3.418 $\times 10^{-5}$ & 44454 & 7597 \\ \hline
	\end{tabular}
		\centering
	\caption{\label{frequency_cavity_modes} Electromagnetic properties of the cavity  modes of at cryogenic temperature. The $Q_0$ was also computed for room temperature using the dimensions calculated taking into account thermal expansion.} 
\end{table}

\begin{figure}[h!]
\begin{center}
\includegraphics[width=1 \textwidth]{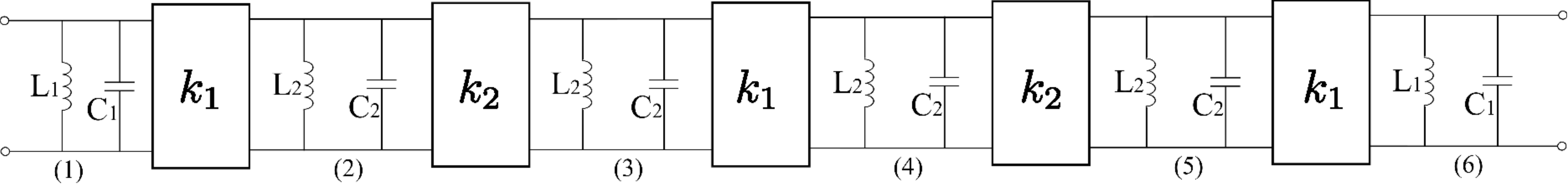}
	\caption{\label{circuit_filter} 
Ideal equivalent circuit used for the synthesis of the six cavities filter.}
\end{center}
\end{figure}

\begin{figure}[h!]
\begin{center}
\includegraphics[width=0.8\textwidth]{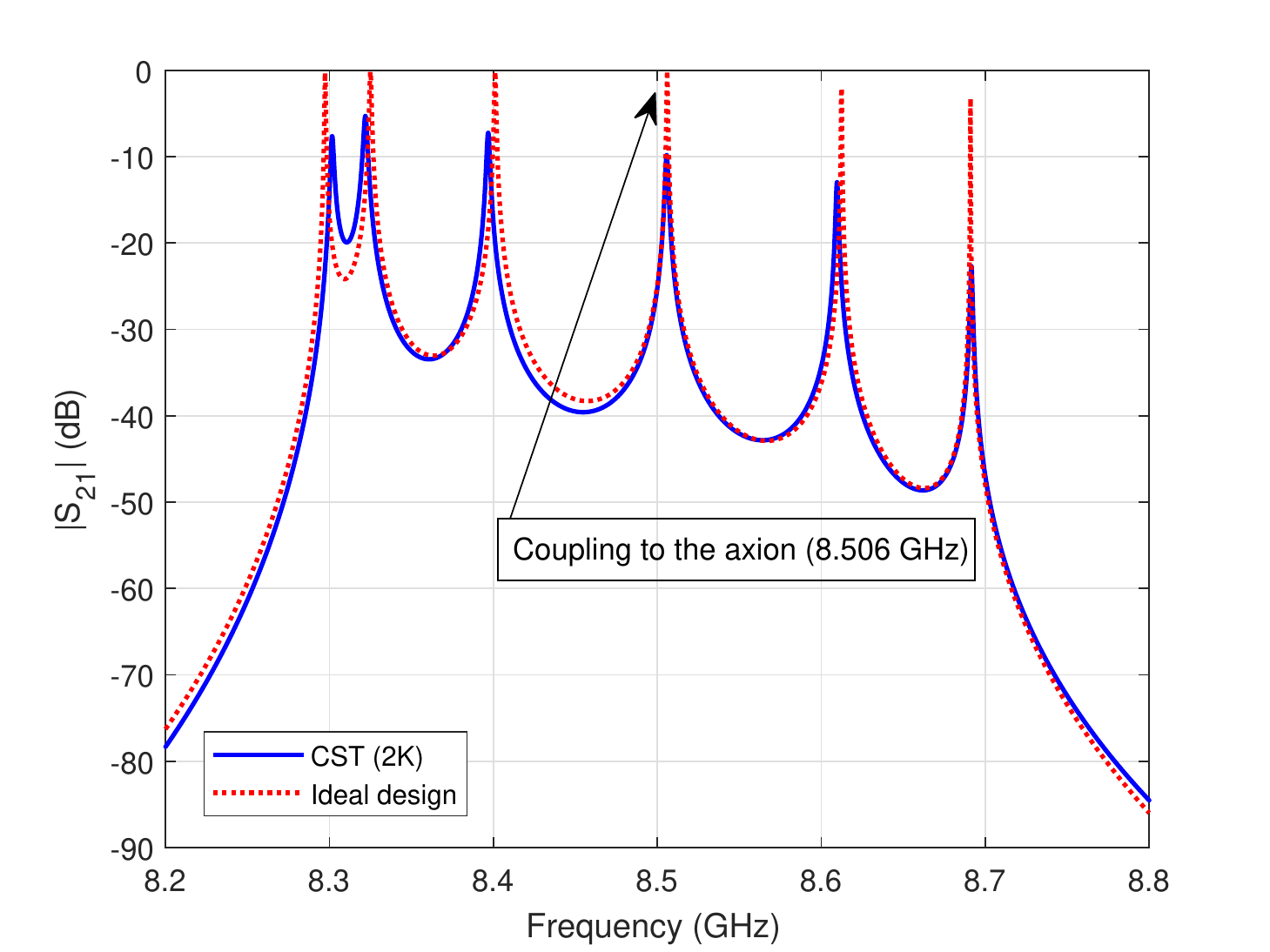}
	\caption{\label{S21_simulation_cryo} 
Transmission coefficient as a function of frequency under cryogenic conditions for a cavity structure simulated with CST Microwave Studio (solid line) compared to the ideal electrical response of the filter (dotted line).}
\end{center}
\end{figure}

In Figure~\ref{circuit_filter} we have represented the ideal network used for the synthesis of the filter \cite{cameron}, which replicates the theoretical model introduced in subsection \ref{s:electro_design}. In this figure $L_q$ and $C_q$ represent the resonators of the structure and determine the $\Omega_q$ of the M-matrix (\ref{OMt}); the elements $k_q$ represent the coupling between adjacent resonators. In this ideal circuit ohmic losses have not been included, and small couplings are used at the input/output cavities following the procedure in \cite{cameron} in order to model the small couplings introduced by the real coaxial connectors. In Table~\ref{lumped_elements} we present the values of the lumped elements of this network. The electrical response (transmission coefficient as a function of frequency) of this ideal network is shown in Fig.~\ref{S21_simulation_cryo} together with the result of a CST simulation. In that simulation ohmic losses have been including using a value of the electrical conductivity\footnote{In this simulation the anomalous skin effect has been neglected which should lead to a slight decrease of $Q_0$ at cryogenic temperatures. With cavities coated with the same procedure we have inferred values closer to  $\sigma_{\unit[2]{K}} \simeq \unit[1 \times 10^9]{S/m}$} $\sigma_{\unit[2]{K}} = \unit[2.008 \times 10^9]{S/m}$. 
This figure enables the comparison of the ideal electrical response (from the mathematical model) with that of the cavity model. We can see that the agreement between the response of the circuit and the full wave simulations is very good.
The fourth resonance shown in this plot corresponds to the working frequency selected for dark matter axion detection. 
\begin{table}[]
	\begin{tabular}{|c|c|}
		\hline
		Component  & Value    \\ \hline
		$L_1$  &  $45.89$ nH        \\ \hline
		$C_1$  &  $7.82$ fF     \\ \hline
		$k_1$  &  $+0.0248$      \\ \hline
		$L_2$  &  $42.84$ nH    \\ \hline
		$C_2$  &  $8.17$ fF             \\ \hline
		$k_2$  &  $-0.0248$ 	 \\ \hline
	\end{tabular}
		\centering
	\caption{\label{lumped_elements} Values of the lumped elements of the ideal circuit used for the synthesis of the filter.} 
\end{table}

\subsection{Tolerances study}
As seen in section \ref{sec:alternating_theory} the alternating design allows the axion signal to be coupled to the central mode. As discussed in section \ref{s:theo_remarks}, the resonant frequency and geometrical factor for this mode are less sensitive to small variations of the geometrical dimensions.

\begin{figure}[h!]
\begin{center}
\includegraphics[width=0.75 \textwidth]{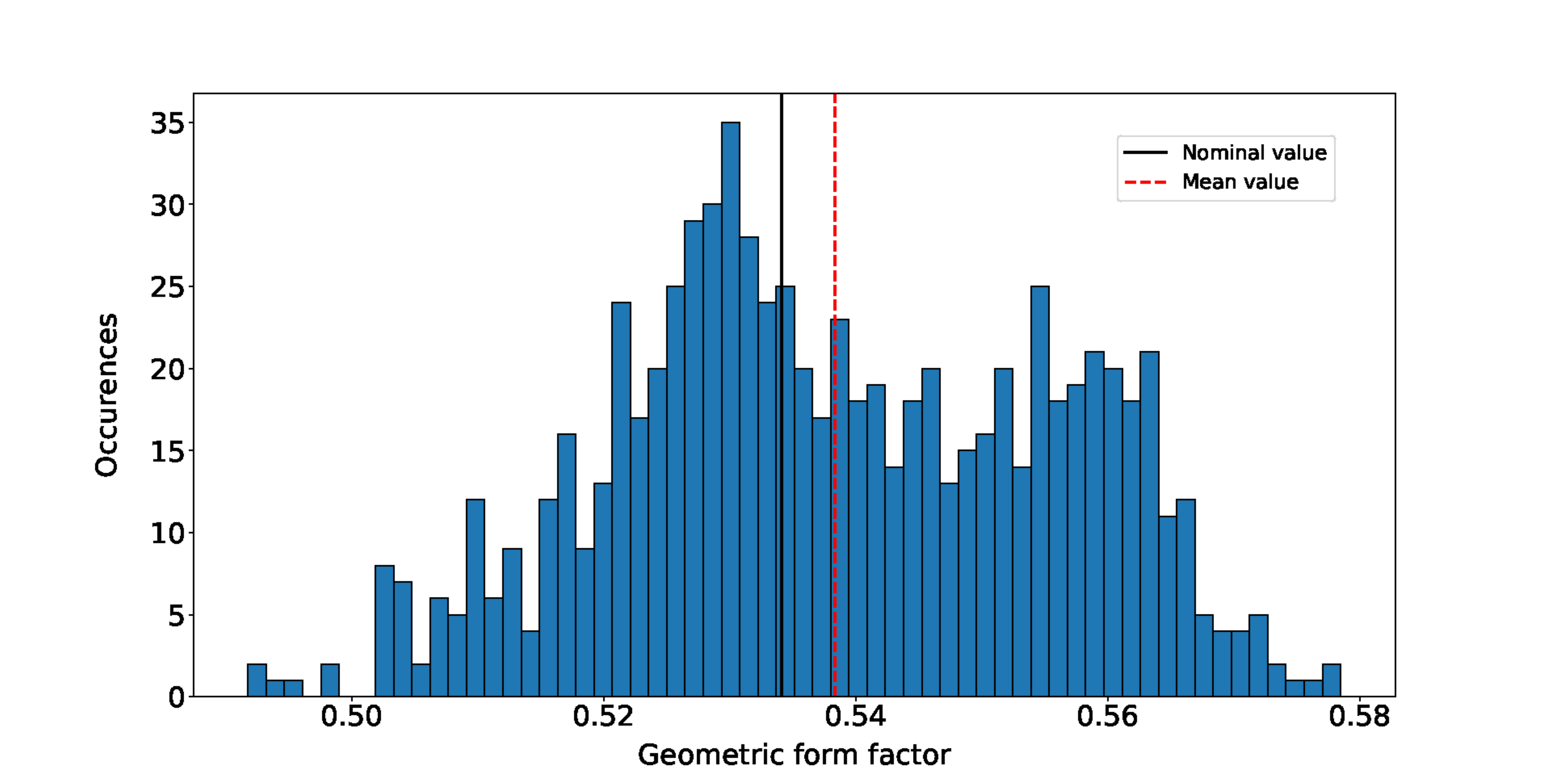}
\caption{Histogram of the result of the tolerance study for the alternating structure. The geometric form factor has been computed $800$ times varying randomly all geometrical dimensions in the range of $\pm 30 \, \mu$m. The nominal and distribution mean values are $0.5340$ and $0.5383$, respectively.}
\label{Tolerances_Histogram}
\end{center}
\end{figure}

To validate the stability of the alternating structure and estimate the performance of a manufactured prototype, a tolerance study was made. For this study, the dimensions of all geometrical parameters (e.g. a, b, l, g, h, w, t) for the different cavities and irises were changed randomly, according to a normal distribution, in the range of $\pm 30 \, \mu$m. With these specifications, we performed $800$ simulations using CST Microwave Studio and calculated for each of them the geometric form factor. The results are summarised in a histogram in Figure~\ref{Tolerances_Histogram} from which we can conclude that the structure is rather robust against mechanical tolerances since, even in the worst case scenario, the geometric form factor drops by less than $10 \%$ with respect to its nominal value.

\subsection{Fabrication of a prototype and characterization}
\label{s:Fabrication}
\begin{figure}[h!]
\begin{center}
\includegraphics[width=0.7 \textwidth]{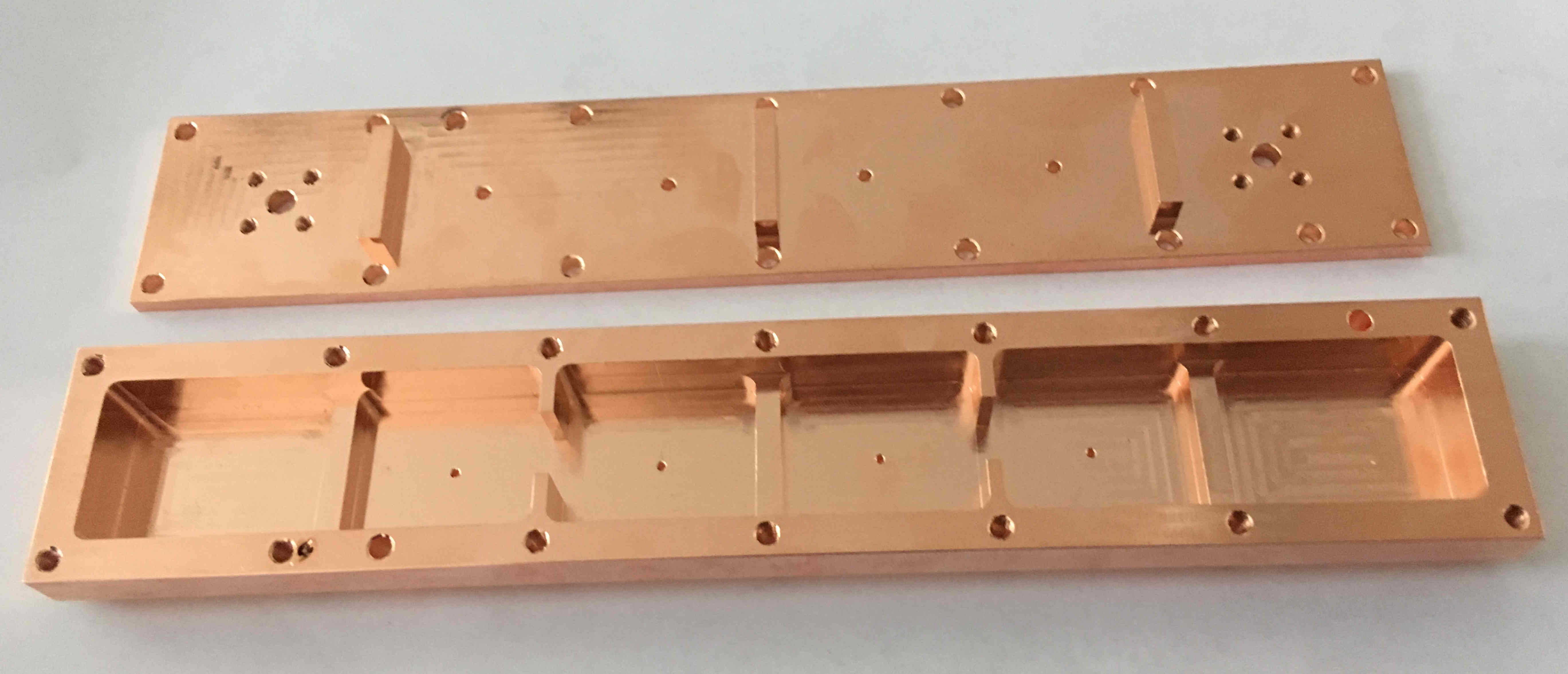}
\caption{\label{photo_filter} Photo of the opened copper-coated 6-cavities alternating prototype. The larger holes at
the outermost edges of the top-plate are used for the placement of the RF antennas.}
\end{center}
\end{figure}

We manufactured our prototype from $316$ LN stainless steel, which is one of the highest quality austenitic steel. The filter was manufactured at Institute Itaca of the Technical University of Valencia \cite{itaca}, and the $\unit[30]{\mu m}$ coating copper layer was applied at CERN. 
A picture of the coated filter is shown in Figure~\ref{photo_filter}.

\begin{figure}[h!]
\begin{center}
\includegraphics[width=0.7 \textwidth]{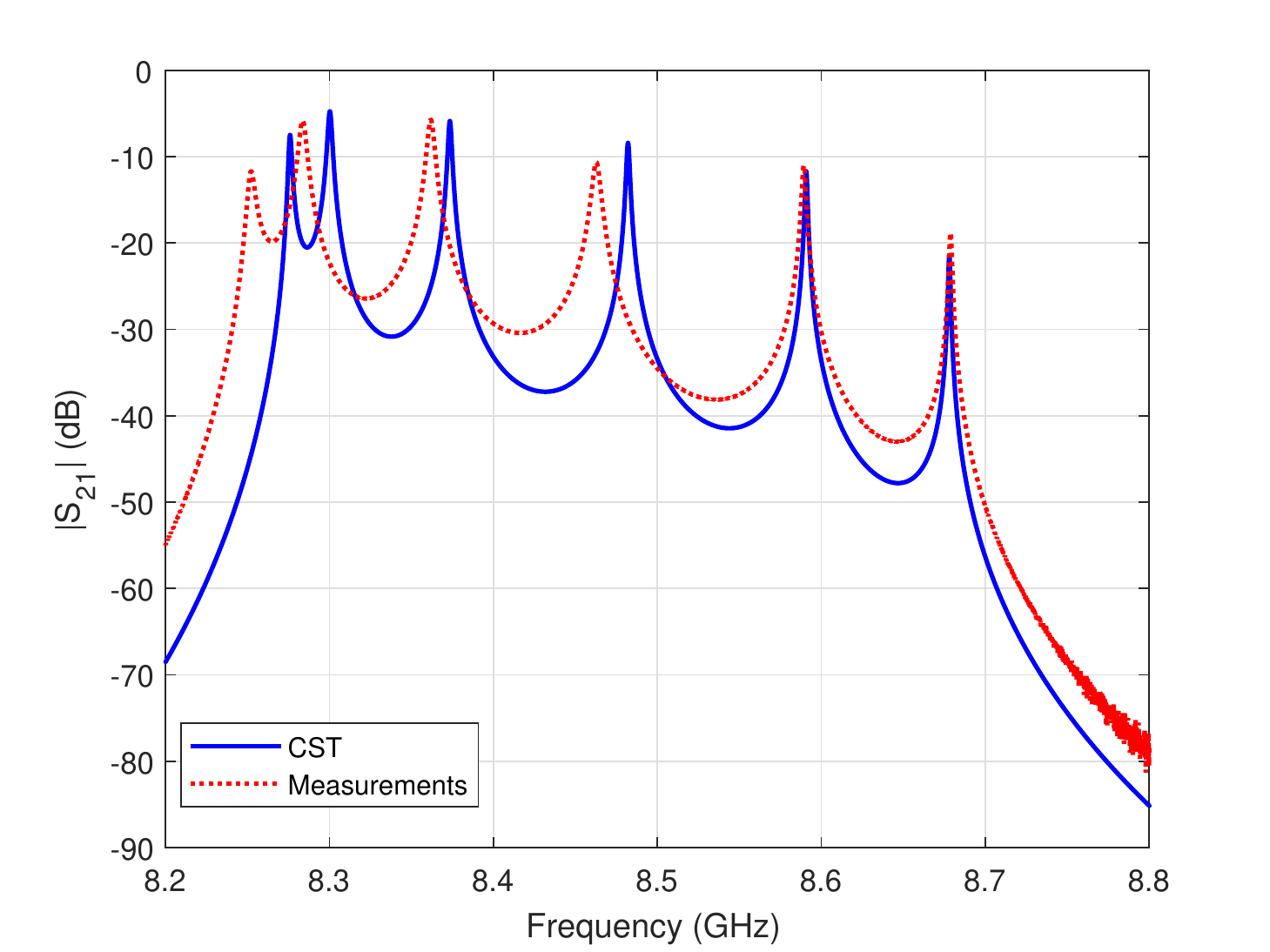}
\caption{\label{S21_exp_CST} Electrical response at ambient temperature of the six cavities filter. Measurement performed with a standard Vector Network Analyzer (dotted line) are compared to CST simulations (solid line).}
\end{center}
\end{figure}

The measured transmission coefficient of the constructed filter at room temperature has been plotted in Figure~\ref{S21_exp_CST} together with the results of the CST simulation using the measured geometrical dimensions (last column of Table~\ref{physical_dimensions}) and electrical conductivity of copper at room temperature.
The discrepancies observed are, we believe, mainly due to the poor contact between the cover and the filter body (see Figure \ref{photo_filter}) caused by the stiffness of the stainless steel employed for manufacturing the filter. Thus, for forthcoming cavities
we are planning to investigate different manufacturing cuts or the possibility of brazing the cavity.

\begin{figure}[h!]
\begin{center}
\includegraphics[width=0.8 \textwidth]{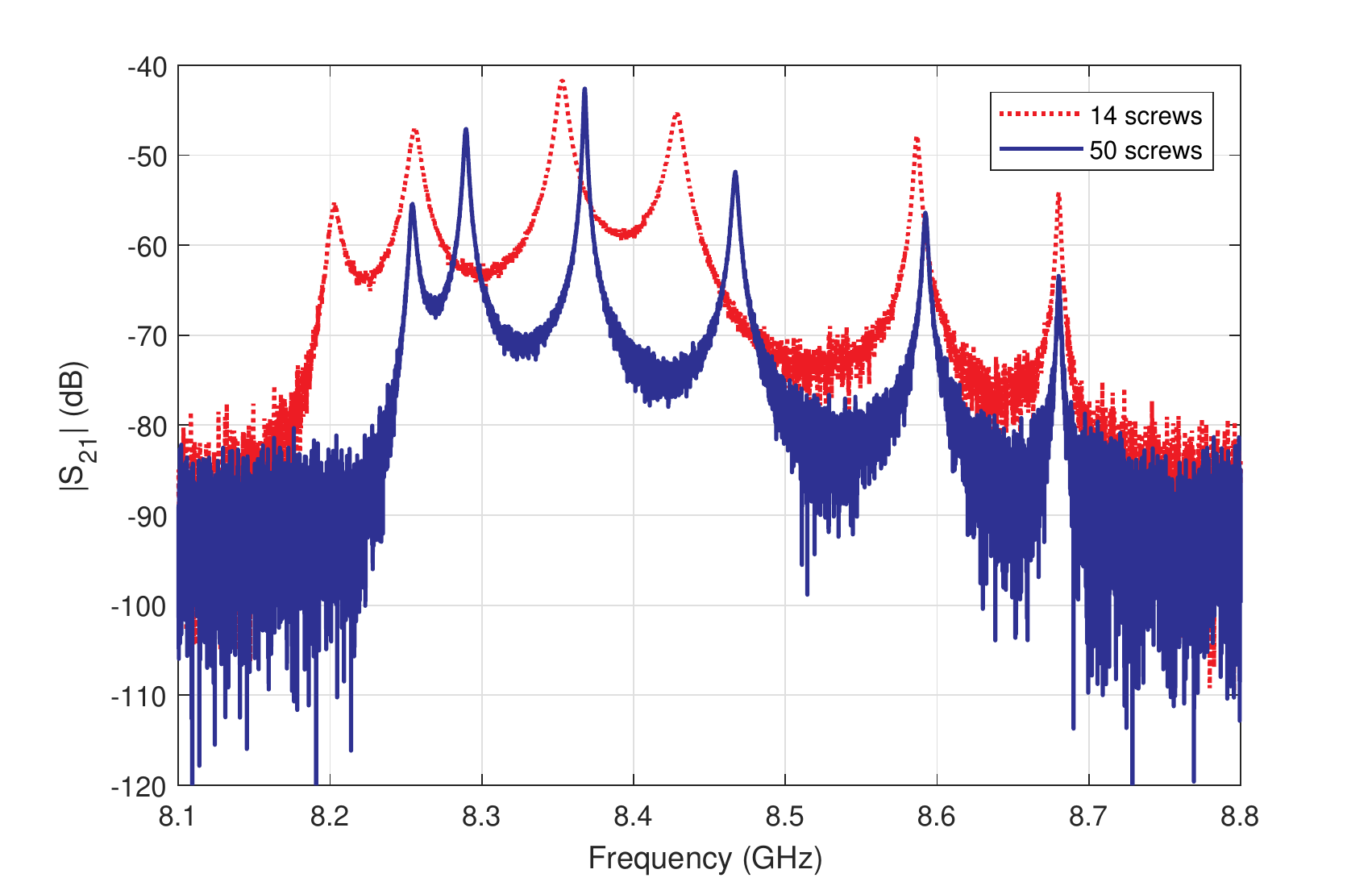}
\caption{\label{morescrews} Electrical response of the six cavities filter for a different number of screws connecting filter and cover. Comparison of the response with 14 screws (dotted line) and 50 screws (solid line). Measurement performed with a standard Vector Network Analyzer at ambient temperature.}
\end{center}
\end{figure}

The determination of the quality factor of the cavity is fundamental to estimate the sensitivity for dark matter axion detection. The well known $-3$ dB method \cite{matthaei} allows a simple calculation of the loaded quality factor $Q_L$ using this expression,
\begin{equation}
    \label{Q-Value} 
    Q_L = \frac{f_0}{\Delta f},
\end{equation}
where $f_0$ is the resonant frequency, and $\Delta f$ is the frequency bandwidth at $\unit[-3]{dB}$ from the peak. This method has been applied to the calculation of $Q_L$ at room temperature of the fourth resonance as depicted in Figure~\ref{S21_exp_CST} (measurement), obtaining $Q_L = 1625$. The unloaded $Q_0$ and external $Q_{ext}$ quality factors are related to $Q_L$ as follows
\begin{equation}
    \label{QL_Q0_Qext} 
    \frac{1}{Q_L} = \frac{1}{Q_0} + \frac{1}{Q_{ext}} .
\end{equation}
The external quality factor $Q_{ext}$ can be obtained using a procedure detailed in \cite{khanna1983} and \cite{bray2004} from the measurement of the loaded quality factor $Q_L$ and the input coupling coefficient yielding $Q_{ext}=5318$ and thus, using Eq.~\ref{QL_Q0_Qext}, $Q_0=2340$. This value has to be compared to the simulated one using room-temperature dimensions and electrical conductivity given in Table \ref{frequency_cavity_modes}. As mentioned before, we consider the factor $\sim3$ difference to be caused by a poor contact between the two parts of the cavity structure. 
Indeed, in a first set of measurements of the filter the discrepancy was even larger ($Q_0=1620$ at ambient temperature). However, adding more screws to improve the mechanical contact between cover and filter showed an increase of the quality factor to $Q_0=2340$ and a frequency shift to higher frequencies of the first four peaks (see Figure \ref{morescrews}), a similar shift as observed between measured and simulated cavity response in Figure~\ref{S21_exp_CST}. Hence, the measurements showed that improving the mechanical contact reduces the discrepancies between theoretical and measured results. In the present setup, the contact could not be made better but we are are investigating measures to mitigate this issue in the future.

\section{Future directions, prospects and conclusions \label{sec:conclusions}}

\begin{figure}[h!]
\begin{center}
\includegraphics[width=1.0 \textwidth]{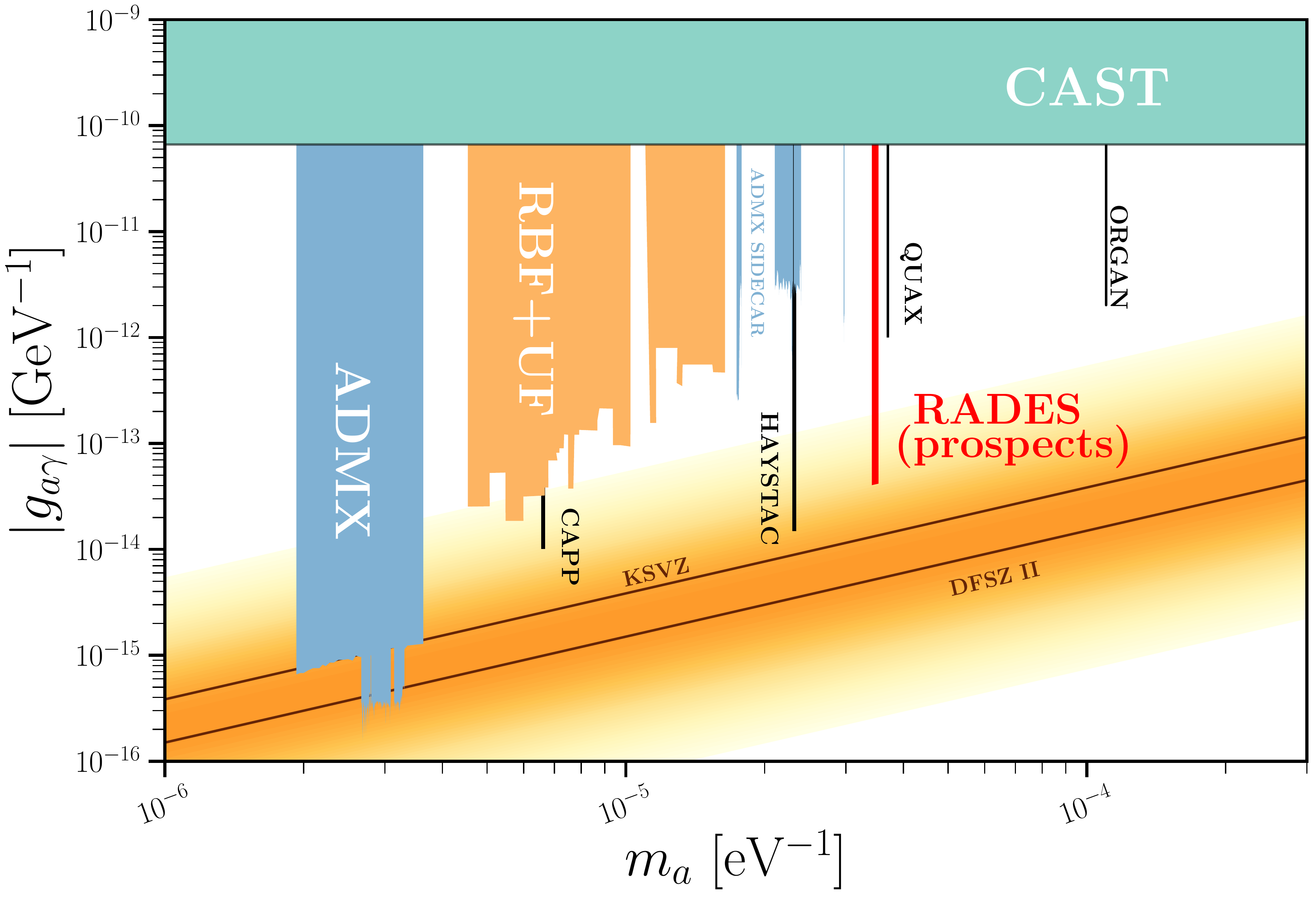}
\caption{\label{prosp} Prospects of the RADES setup presented in this work in the context of other haloscope searches: BFRT-UF~\cite{DePanfilis:1987dk,Hagmann:1990tj}, ADMX~\cite{Du:2018uak, Braine:2019fqb}, HAYSTAC~\cite{Zhong:2018rsr}, QUAX~\cite{Alesini:2019ajt} and ORGAN~\cite{McAllister:2017lkb}.
        }
\end{center}
\end{figure}

\begin{figure}[h!]
\begin{center}
\includegraphics[width=0.5 \textwidth]{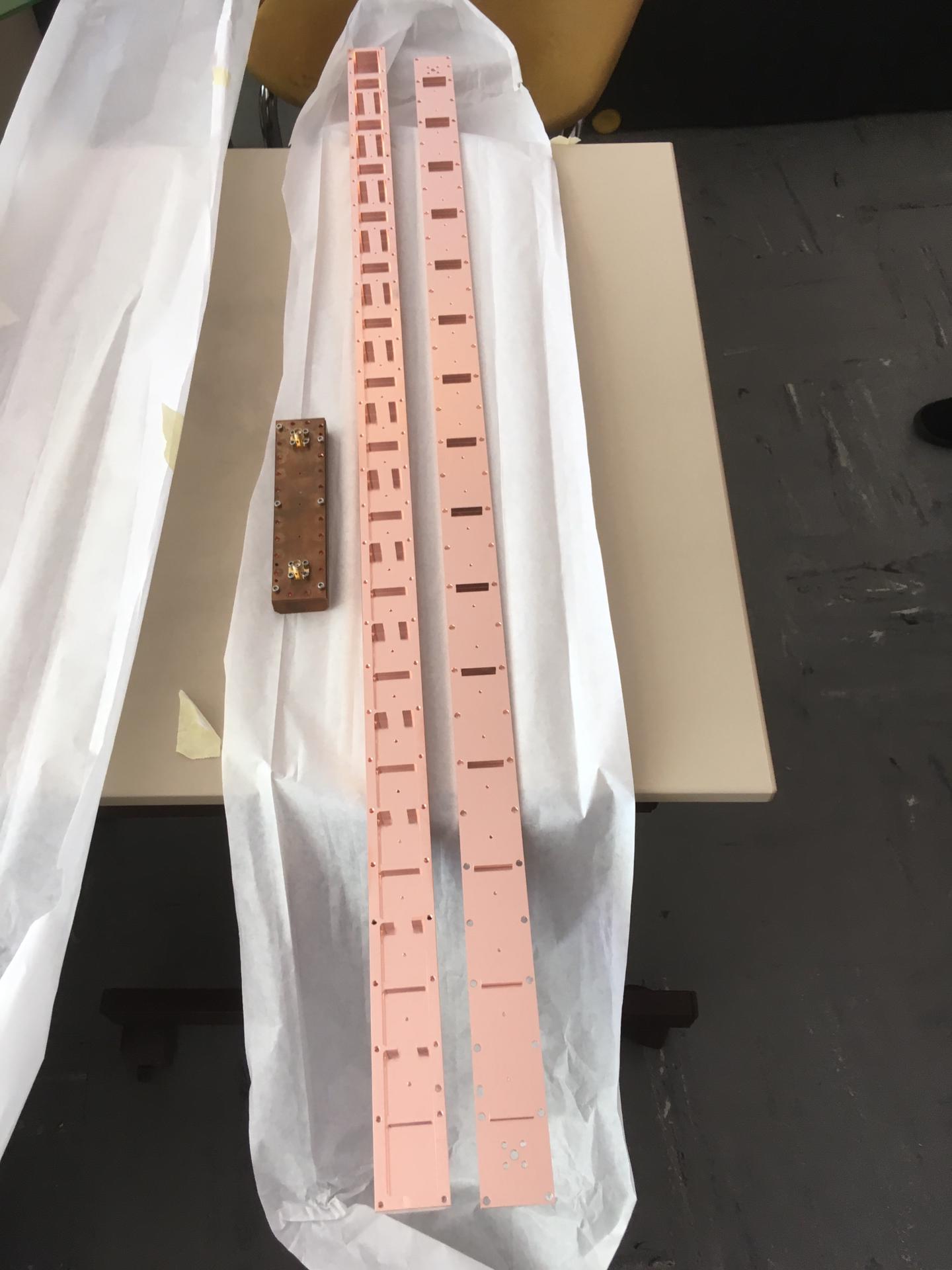}
\caption{\label{1m_cavity} Comparison of the fully inductive cavity reported in \cite{Melcon:2018dba} (left) with the $\sim$ 1 meter long cavity (right) based on the new concept put forward in this paper.}
\end{center}
\end{figure}

In this article we have discussed the concept and realisation of a new cavity design useful in the search for $>\unit[30]{\mu eV}$ axions with microwave filters. As such filters can be used in any dipole magnetic field, this is of general interest for `axion hunters' all over the world and
a prime example of a `physics beyond colliders' initiative \cite{Alemany:2019vsk},
combining accelerator technology (microwave filters and dipole magnets) at CERN
for particle search outside colliders.
Our developments constitute a particularly important step towards searching axions at relatively large masses with the future International Axion Observatory (IAXO). We have theoretically shown that a structure of alternating capacitive and inductive irises couples to the axion at a higher order resonance, alleviating the problem of mode-mixing of previous designs. We have also built and experimentally characterised a prototype that demonstrated the expected good behaviour of this structure in practice. The proposed improvements potentially allow to produce much longer cavities for axion search than previously reported prototypes.
In fact, at the time of writing, a $\sim$ 1 meter long cavity of this type is installed in the CAST magnet and is foreseen to take physics data in 2020. A photograph of this longer manufactured prototype is presented in Figure~\ref{1m_cavity}, together with a smaller-sized cavity with similar dimensions as the one described in this paper.
The longer prototype was also produced with the top plate technique, and thus we observed similar issues with the
quality factor as discussed in the end of the previous section.
Despite this shortcoming on which we are working, we briefly discuss here the
prospects to search for axions within the CAST magnet
with this alternating-type structure of greater volume.
We find that we can explore axion couplings down to $g_{A\gamma}\sim 10^{-14}$ GeV$^{-1}$ at $m\simeq 35 \mu$eV within a few months of data in the CAST magnet, see Figure \ref{prosp} .
Expressing the coupling as \cite{Irastorza:2018dyq}
\begin{equation}
g_{A\gamma} \equiv 2.0 \times 10^{-16} C_{A \gamma} \frac{m_a}{\mu {\rm eV}} {\rm GeV}^{-1} \ ,
\end{equation}
we can compute our sensitivity through\footnote{Thereby correcting a small typo in \cite{Melcon:2018dba}. }

\begin{eqnarray}
\left.C_{A \gamma}\right|_{\rm reach}
&\simeq & 26.1
\left(\frac{\frac{S}{N}}{3}\right)^\frac{1}{2}\frac{9\, \rm T}{B_e}\left(\frac{1\, \rm l}{V}\right)^\frac{1}{2}
\left(\frac{10^4}{Q}\right)^\frac{1}{2} \left(\frac{0.69}{{{\mathcal{G}_i}}}\right)
\left(\frac{T_{\rm eff}}{10\, \rm K}\right)^\frac{1}{2}
\left(\frac{0.5}{\kappa}\right)^\frac{1}{2}
\left(\frac{30 \rm \mu eV}{m_a}\frac{\rm hour}{t}\right)^\frac{1}{4} \nonumber .
\label{eq:sensi}
\end{eqnarray}
In Fig.~\ref{prosp}  we set $Q=1200$ and coupling $\kappa$= 0.43  (as measured in situ), a central frequency of $\sim$8.4 GHz and thus a mass of m = 34.74 $\mu$eV, a geometrical factor of $\mathcal{G}=0.55$, a noise temperature of 10K as well as a data taking time of around 6 months. 
To put into context the relevance of this measurement, we also include existing limits from the literature. It is visible that RADES can compete with the major players in axion search, such as the `high-mass ADMX’-project `SIDECAR' whilst tapping into a completely unexplored mass reach in between SIDECAR and QUAX.

An important aspect for future study will be how to tune this type of cavity structures.
While tuning is solved in principle for the inductive-cavity type, see \cite{Cuendis:2019qij}, a mechanical tuning for the alternating structure still has to be developed. For this reason, we are investigating possibilities to achieve tuning 
with ferroelectric materials \cite{PATRASBarcelo:2019}. Physics results using either technique will be described in forthcoming articles.
%

\section*{Acknowledgements}

We thank Ciaran O'Hare for his generous and publicly available compilation of axion bounds \url{https://github.com/cajohare/AxionLimits/}. 
This work has been funded by the Spanish Ministerio de Economía, Industria y Competitividad – Agencia Estatal de Investigacion (AEI) and Fondo Europeo de Desarrollo Regional (FEDER) under project FPA-2016-76978, and was supported by the CERN Doctoral Studentship programme. The research leading to these results has received funding from the European Research Council and BD, JG and SAC acknowledge support through the European Research Council under grant ERC-2018-StG-802836  (AxScale project). IGI acknowledges also support from the European Research Council (ERC) under grant ERC-2017-AdG-788781 (IAXO+ project). JR has been supported by the Ramon y Cajal Fellowship 2012-10597, the grant PGC2018-095328-B-I00(FEDER\slash Agencia  estatal  de  investigaci\'on)  and  FSE-DGA2017-2019-E12/7R (Gobierno de Arag\'on/FEDER) (MINECO\slash FEDER), the EU through the ITN “Elusives” H2020-MSCA-ITN-2015\slash674896 and the Deutsche Forschungsgemeinschaft under grant SFB-1258 as a Mercator Fellow. CPG was supported by PROMETEO II\slash2014\slash050 of Generalitat Valenciana, FPA2014-57816-P of MINECO and by the European Union’s Horizon 2020 research and innovation program under the Marie Sklodowska-Curie grant agreements 690575 and 674896. AM is supported by the
European Research Council under Grant No. 742104.
 We wish also to thank our colleagues at CAST and at CERN,  in particular
 Marc Thiebert from the coating lab,
 Sergio Calatroni for many useful discussions,
 as well as the whole team of the CERN Central Cryogenic Laboratory for their support and advice in specific aspects of the project.


\end{document}